\errorstopmode
\input amssym.def
\input amssym.tex


\magnification=\magstephalf
\hsize=14.0 true cm
\vsize=19 true cm
\hoffset=1.0 true cm
\voffset=2.0 true cm

\abovedisplayskip=12pt plus 3pt minus 3pt
\belowdisplayskip=12pt plus 3pt minus 3pt
\parindent=1.0em


\font\sixrm=cmr6
\font\eightrm=cmr8
\font\ninerm=cmr9

\font\sixi=cmmi6
\font\eighti=cmmi8
\font\ninei=cmmi9

\font\sixsy=cmsy6
\font\eightsy=cmsy8
\font\ninesy=cmsy9

\font\sixbf=cmbx6
\font\eightbf=cmbx8
\font\ninebf=cmbx9

\font\eightit=cmti8
\font\nineit=cmti9

\font\eightsl=cmsl8
\font\ninesl=cmsl9

\font\sixss=cmss8 at 8 true pt
\font\sevenss=cmss9 at 9 true pt
\font\eightss=cmss8
\font\niness=cmss9
\font\tenss=cmss10

\font\sixmib=cmmib6
\font\sevenmib=cmmib7
\font\eightmib=cmmib8
\font\ninemib=cmmib9
\font\tenmib=cmmib10

 at 12 true pt
 at 12 true pt
\font\bigrm=cmr10 at 12 true pt
 at 12 true pt
 at 12 true pt

 at 16 true pt
 at 16 true pt
\font\Bigrm=cmr12 at 16 true pt
 at 16 true pt
 at 16 true pt

\catcode`@=11
\newfam\ssfam
\newfam\mibfam

\def\tenpoint{\def\rm{\fam0\tenrm}%
    \textfont0=\tenrm \scriptfont0=\sevenrm \scriptscriptfont0=\fiverm
    \textfont1=\teni  \scriptfont1=\seveni  \scriptscriptfont1=\fivei
    \textfont2=\tensy \scriptfont2=\sevensy \scriptscriptfont2=\fivesy
    \textfont3=\tenex \scriptfont3=\tenex   \scriptscriptfont3=\tenex
    \textfont\itfam=\tenit                  \def\it{\fam\itfam\tenit}%
    \textfont\slfam=\tensl                  \def\sl{\fam\slfam\tensl}%
    \textfont\bffam=\tenbf \scriptfont\bffam=\sevenbf
                           \scriptscriptfont\bffam=\fivebf
                           \def\bf{\fam\bffam\tenbf}%
    \textfont\ssfam=\tenss \scriptfont\ssfam=\sevenss
                           \scriptscriptfont\ssfam=\sevenss
                           \def\ss{\fam\ssfam\tenss}%
    \textfont\mibfam=\tenmib \scriptfont\mibfam=\sevenmib
                             \scriptscriptfont\mibfam=\sevenmib
                             \def\mib{\fam\mibfam\tenmib}%
    \normalbaselineskip=13pt
    \setbox\strutbox=\hbox{\vrule height8.5pt depth3.5pt width0pt}%
    \let\big=\tenbig
    \normalbaselines\rm}

\def\ninepoint{\def\rm{\fam0\ninerm}%
    \textfont0=\ninerm      \scriptfont0=\sixrm
                            \scriptscriptfont0=\fiverm
    \textfont1=\ninei       \scriptfont1=\sixi
                            \scriptscriptfont1=\fivei
    \textfont2=\ninesy      \scriptfont2=\sixsy
                            \scriptscriptfont2=\fivesy
    \textfont3=\tenex       \scriptfont3=\tenex
                            \scriptscriptfont3=\tenex
    \textfont\itfam=\nineit \def\it{\fam\itfam\nineit}%
    \textfont\slfam=\ninesl \def\sl{\fam\slfam\ninesl}%
    \textfont\bffam=\ninebf \scriptfont\bffam=\sixbf
                            \scriptscriptfont\bffam=\fivebf
                            \def\bf{\fam\bffam\ninebf}%
    \textfont\ssfam=\niness \scriptfont\ssfam=\sixss
                            \scriptscriptfont\ssfam=\sixss
                            \def\ss{\fam\ssfam\niness}%
    \textfont\mibfam=\ninemib \scriptfont\mibfam=\sixmib
                            \scriptscriptfont\mibfam=\sixmib
                            \def\mib{\fam\mibfam\ninemib}%
    \normalbaselineskip=12pt
    \setbox\strutbox=\hbox{\vrule height8.0pt depth3.0pt width0pt}%
    \let\big=\ninebig
    \normalbaselines\rm}

\def\eightpoint{\def\rm{\fam0\eightrm}%
    \textfont0=\eightrm      \scriptfont0=\sixrm
                             \scriptscriptfont0=\fiverm
    \textfont1=\eighti       \scriptfont1=\sixi
                             \scriptscriptfont1=\fivei
    \textfont2=\eightsy      \scriptfont2=\sixsy
                             \scriptscriptfont2=\fivesy
    \textfont3=\tenex        \scriptfont3=\tenex
                             \scriptscriptfont3=\tenex
    \textfont\itfam=\eightit \def\it{\fam\itfam\eightit}%
    \textfont\slfam=\eightsl \def\sl{\fam\slfam\eightsl}%
    \textfont\bffam=\eightbf \scriptfont\bffam=\sixbf
                             \scriptscriptfont\bffam=\fivebf
                             \def\bf{\fam\bffam\eightbf}%
    \textfont\ssfam=\eightss \scriptfont\ssfam=\sixss
                             \scriptscriptfont\ssfam=\sixss
                             \def\ss{\fam\ssfam\eightss}%
    \textfont\mibfam=\eightmib \scriptfont\mibfam=\sixmib
                             \scriptscriptfont\mibfam=\sixmib
                             \def\mib{\fam\mibfam\eightmib}%
    \normalbaselineskip=10pt
    \setbox\strutbox=\hbox{\vrule height7.0pt depth2.0pt width0pt}%
    \let\big=\eightbig
    \normalbaselines\rm}

\def\tenbig#1{{\hbox{$\left#1\vbox to8.5pt{}\right.\n@space$}}}
\def\ninebig#1{{\hbox{$\textfont0=\tenrm\textfont2=\tensy
                       \left#1\vbox to7.25pt{}\right.\n@space$}}}
\def\eightbig#1{{\hbox{$\textfont0=\ninerm\textfont2=\ninesy
                       \left#1\vbox to6.5pt{}\right.\n@space$}}}

\font\sectionfont=cmbx10
\font\subsectionfont=cmti10

\def\figurecaptionfont{\ninepoint}
\def\tablecaptionfont{\ninepoint}
\def\footnotefont{\eightpoint}


\newcount\equationno
\newcount\bibitemno
\newcount\figureno
\newcount\tableno

\equationno=0
\bibitemno=0
\figureno=0
\tableno=0


\footline={\ifnum\pageno=0{\hfil}\else
{\hss\rm\the\pageno\hss}\fi}


\def\section #1. #2 \par
{\vskip0pt plus .10\vsize\penalty-100 \vskip0pt plus-.10\vsize
\vskip 1.6 true cm plus 0.2 true cm minus 0.2 true cm
\global\def\equationlabel{#1}
\global\equationno=0
\leftline{\sectionfont #1. #2}\par
\immediate\write\terminal{Section #1. #2}
\vskip 0.7 true cm plus 0.1 true cm minus 0.1 true cm
\noindent}


\def\subsection #1 \par
{\vskip0pt plus 1.0 true cm\penalty-50 \vskip0pt plus-1.0 true cm
\vskip2.5ex plus 0.1ex minus 0.1ex
\leftline{\subsectionfont #1}\par
\immediate\write\terminal{Subsection #1}
\vskip1.0ex plus 0.1ex minus 0.1ex
\noindent}


\def\appendix #1. #2 \par
{\vskip0pt plus .10\vsize\penalty-100 \vskip0pt plus-.10\vsize
\vskip 1.6 true cm plus 0.2 true cm minus 0.2 true cm
\global\def\equationlabel{\hbox{\rm#1}}
\global\equationno=0
\leftline{\sectionfont Appendix #1. #2}\par
\immediate\write\terminal{Appendix #1. #2}
\vskip 0.7 true cm plus 0.1 true cm minus 0.1 true cm
\noindent}



\def\equation#1{$$\displaylines{\qquad #1}$$}
\def\enum{\global\advance\equationno by 1
\hfill\llap{{\rm(\equationlabel.\the\equationno)}}}
\def\noenum{\hfill}

\def\nexteq#1{\cr\noalign{\vskip#1}\qquad}


\def\ifundefined#1{\expandafter\ifx\csname#1\endcsname\relax}

\def\ref#1{\ifundefined{#1}?\immediate\write\terminal{unknown reference
on page \the\pageno}\else\csname#1\endcsname\fi}

\newwrite\terminal
\newwrite\bibitemlist

\def\bibitem#1#2\par{\global\advance\bibitemno by 1
\immediate\write\bibitemlist{\string\def
\expandafter\string\csname#1\endcsname
{\the\bibitemno}}
\item{[\the\bibitemno]}#2\par}

\def\beginbibliography{
\vskip0pt plus .15\vsize\penalty-100 \vskip0pt plus-.15\vsize
\vskip 1.2 true cm plus 0.2 true cm minus 0.2 true cm
\leftline{\sectionfont References}\par
\immediate\write\terminal{References}
\immediate\openout\bibitemlist=biblist
\frenchspacing\parindent=1.8em
\vskip 0.5 true cm plus 0.1 true cm minus 0.1 true cm}

\def\endbibliography{
\immediate\closeout\bibitemlist
\nonfrenchspacing\parindent=1.0em}


\def\figurecaption#1{\global\advance\figureno by 1
\narrower\figurecaptionfont Fig.~\the\figureno. #1}

\def\tablecaption#1{\global\advance\tableno by 1
\hfill{\tablecaptionfont Table~\the\tableno. #1}\hfill}

\def\thicktablerule{\hrule height0.8pt}
\def\thintablerule{\hrule height0.4pt}

\tenpoint

\immediate\openin\bibitemlist=biblist
\ifeof\bibitemlist\immediate\closein\bibitemlist
\else\immediate\closein\bibitemlist

 \fi


\def\thismonth{\ifcase\month\or
January\or February\or March\or April\or May\or June\or
July\or August\or September\or October\or November\or December\fi}

\input epsf
\epsfclipon



\def\rmd{{\rm d}}

\def\rme{{\rm e}}
\def\rmO{{\rm O}}


\def\Re{{\rm Re}\,}


\def\proof{\noindent{\sl Proof:}\kern0.6em}

\def\frac#1#2{\hbox{$#1\over#2$}}
\def\dual{\mathstrut^*\kern-0.1em}

\def\ring{\mathaccent"7017}
\def\lvec#1{\setbox0=\hbox{$#1$}
    \setbox1=\hbox{$\scriptstyle\leftarrow$}
    #1\kern-\wd0\smash{
    \raise\ht0\hbox{$\raise1pt\hbox{$\scriptstyle\leftarrow$}$}}
    \kern-\wd1\kern\wd0}
\def\rvec#1{\setbox0=\hbox{$#1$}
    \setbox1=\hbox{$\scriptstyle\rightarrow$}
    #1\kern-\wd0\smash{
    \raise\ht0\hbox{$\raise1pt\hbox{$\scriptstyle\rightarrow$}$}}
    \kern-\wd1\kern\wd0}

\def\slash#1{\setbox2=\hbox{$\displaystyle#1$}%
             \setbox3=\hbox{$\displaystyle/$}%
             #1\kern-0.8\wd2/\kern-1.0\wd3\kern0.8\wd2\kern0.5pt}
\def\ttilde#1{\setbox4=\hbox{$\displaystyle\tilde{\phantom{#1}}$}%
                    \hbox{$\displaystyle\tilde{\phantom{#1}}$}\kern-1.0\wd4%
         \raise1.6pt\hbox{$\displaystyle\tilde{\phantom{#1}}$}\kern-1.0\wd4%
         #1}
\def\sttilde#1{\setbox4=\hbox{$\scriptstyle\tilde{\phantom{#1}}$}%
                    \hbox{$\scriptstyle\tilde{\phantom{#1}}$}\kern-1.0\wd4%
         \raise1.0pt\hbox{$\scriptstyle\tilde{\phantom{#1}}$}\kern-1.0\wd4%
         #1}

\def\wick#1{\setbox2=\hbox{$\displaystyle#1$}
    \setbox3=\null\ht3=3.0pt\dp3=0.0pt\wd3=20.0pt
    #1\kern-\wd2\kern3.0pt\raise11.0pt\vbox{\hrule height0.3pt
    \hbox{\vrule width0.3pt\box3\vrule width0.3pt}}\kern-24.0pt\kern\wd2}

\def\longwick#1{\setbox2=\hbox{$\displaystyle#1$}
    \setbox3=\null\ht3=3.0pt\dp3=0.0pt\wd3=27.0pt
    #1\kern-\wd2\kern3.0pt\raise11.0pt\vbox{\hrule height0.3pt
    \hbox{\vrule width0.3pt\box3\vrule width0.3pt}}\kern-31.0pt\kern\wd2}

\def\verylongwick#1{\setbox2=\hbox{$\displaystyle#1$}
    \setbox3=\null\ht3=3.0pt\dp3=0.0pt\wd3=43.0pt
    #1\kern-\wd2\kern3.0pt\raise11.0pt\vbox{\hrule height0.3pt
    \hbox{\vrule width0.3pt\box3\vrule width0.3pt}}\kern-47.0pt\kern\wd2}


\def\nab#1{{\nabla_{#1}}}
\def\nabstar#1{{\nabla\kern0.5pt\smash{\raise 4.5pt\hbox{$\ast$}}
               \kern-5.5pt_{#1}}}
\def\drv#1{{\partial_{#1}}}
\def\drvstar#1{{\partial\kern0.5pt\smash{\raise 4.5pt\hbox{$\ast$}}
               \kern-6.0pt_{#1}}}

\def\ldrvstar#1{{\lvec{\,\partial}\kern-0.5pt\smash{\raise 4.5pt\hbox{$\ast$}}
               \kern-5.0pt_{#1}}}


\def\MeV{{\rm MeV}}

\def\MSbar{\overline{\rm MS\kern-0.5pt}\kern0.5pt}



\def\psibar{\overline{\psi}}
\def\chibar{\overline{\chi}}
\def\phibar{\overline{\phi}}
\def\lambdabar{\overline{\lambda}}

\def\cbar{\bar{c}}
\def\dbar{\bar{d}}

\def\Aax{A}
\def\Pax{P}
\def\Ptax{\tilde{\Pax}}

\def\ren#1{#1_{\hbox{\sixrm R}}}

\def\imp#1{#1_{\hbox{\sixrm I}}}

\def\rens#1#2{#1_{\hbox{\sixrm R},#2}}
\def\imps#1#2{#1_{\hbox{\sixrm I},#2}}


\def\dirac#1{\gamma_{#1}}
\def\diracstar#1#2{
    \setbox0=\hbox{$\gamma$}\setbox1=\hbox{$\gamma_{#1}$}
    \gamma_{#1}\kern-\wd1\kern\wd0
    \smash{\raise4.5pt\hbox{$\scriptstyle#2$}}}


\def\SUn{{\rm SU}(N)}

\def\tr{{\rm tr}}

\def\Psu{{\cal P}}
\def\Dsu{\frak{N}}


\def\SF{S_{\rm F}}
\def\Sw{S_{\rm w}}
\def\SGfl{S_{\rm G,fl}}
\def\SFfl{S_{\rm F,fl}}
\def\SFPfl{S_{\rm FP,fl}}
\def\Sms{S_{\rm ms}}
\def\Dw{D_{\rm w}}
\def\Dwdag{{\Dw}\kern-4pt^{\dagger}\kern1pt}
\def\csw{c_{\rm sw}}
\def\ZA{Z_A}
\def\ZP{Z_P}
\def\ZAh#1{\hat{Z}_A^{#1}}
\def\ZPh#1{\hat{Z}_P^{#1}}


\def\Dm{D}
\def\Dmdag{\Dm^{\dagger}\kern-1pt}

\def\Dmsdag{D_s\kern-2pt\vphantom{D}^{\dagger}\kern-1pt}


\def\mR#1{m_{\hbox{\sixrm R},#1}}


\def\csw{c_{\rm sw}}
\def\cfl{c_{\rm fl}}
\def\ca{c_A}
\def\cta{\tilde{c}_A}
\def\ctp{\tilde{c}_P}
\def\mq#1{m_{{\rm q},#1}}
\def\Mq{M_{\rm q}}


\def\CPP#1{{\cal C}_{P}^{#1}}
\def\CPtP#1{{\cal C}_{\tilde{P}}^{#1}}
\def\CPhP#1{{\cal C}_{\hat{P}}^{#1}}
\def\CAP#1{{\cal C}_{\partial\kern-0.5pt A}^{#1}}
\def\CAtP#1{{\cal C}_{\partial\kern-0.5pt\tilde{A}}^{#1}}
\def\CdPP#1{{\cal C}_{\partial\partial\kern-0.5pt P}^{#1}}
\def\CQP#1{{\cal C}_{Q}^{#1}}


\def\eps{\epsilon}
\def\Mpi{M_{\pi}}
\def\Fpi{F_{\pi}}
\def\Gpi{G_{\pi}}
\def\Gpit{G_{\pi,t}}

\def\CF{C_{\rm F}}
\def\Sigt{\Sigma_t}

\def\mbar{\kern0.5pt\overline{\kern-0.5pt m\kern-0.8pt}\kern0.8pt}
\def\zbar{\kern0.4pt\overline{\kern-0.4pt z\kern-0.6pt}\kern0.6pt}
\def\Zbar{\kern1.4pt\overline{\kern-1.4pt Z\kern-0.8pt}\kern0.8pt}
\def\Sigbar{\overline{\Sigma\kern-0.8pt}{\vphantom{\Sigma}}\kern0.8pt_t}
\def\cvec#1{\kern-0.5pt\vec{\kern0.5pt #1}}

%
\rightline{CERN-PH-TH/2013-025}
\vskip1.2cm 
\centerline{\Bigrm
Chiral symmetry and the Yang--Mills gradient flow}

\vskip 0.6 true cm
\centerline{\bigrm Martin L\"uscher}
\vskip1.5ex
\centerline{{\it CERN, Physics Department, 1211 Geneva 23, Switzerland}}
\vskip 0.8 true cm
\thintablerule
\vskip 2.0ex
\ninepoint
\leftline{\bf Abstract}
\vskip 1.0ex\noindent
In the last few years,
the Yang--Mills gradient flow was shown to
be an attractive tool for non-perturbative studies
of non-Abelian gauge theories.
Here a simple extension of the 
flow to the quark fields in QCD is considered.
As in the case of the pure-gauge gradient flow,
the renormalizability of correlation functions involving
local fields at positive flow times can be established
using a representation through a local field theory in
$4+1$ dimensions. 
Applications of the extended flow in lattice QCD
include non-perturbative renormalization
and O($a$) improvement as well as
accurate calculations of the chiral condensate 
and of the pseudo-scalar decay constant in the chiral limit.
\vskip 2.0ex
\thintablerule

\tenpoint

\vskip-0.3cm

\section 1. Introduction

In view of its renormalization properties 
[\ref{WilsonFlow},\ref{RenFlow}], and since
its application in lattice gauge theory is technically straightforward,
the Yang--Mills gradient flow allows the dynamics of non-Abelian
gauge theories to be probed in many interesting ways.
The flow can be used for accurate scale setting, for example,
and it provides an understanding of how exactly
the topological (instanton) sectors emerge in
the continuum limit of lattice QCD
[\ref{WilsonFlow}]\kern1pt\footnote{$\dagger$}{\footnotefont%
In lattice gauge theory, where the right-hand side of the flow
equation coincides with the gradient of the Wilson plaquette action,
the flow is often referred to as the ``Wilson flow''. 
The term ``gradient flow'' is
used here both in the continuum theory and on the lattice.}.
Moreover, observables at positive flow time 
are natural quantities to consider for non-perturbative
renormalization and step scaling 
[\ref{StepScaling}--\ref{FritzschRamos}].

Matter fields may or may not be included in the flow equations.
In this paper, a fairly trivial extension of the flow to the 
quark fields in QCD is considered,
where the flow equation for the
gauge field is unchanged, while the evolution of 
the quark fields as a function of the flow time is determined
by a gauge-covariant heat equation (see sect.~2).
The theoretical analysis and practical implementation of the flow 
is not significantly complicated by
the inclusion of the quark fields. In particular,
following ref.~[\ref{RenFlow}], the renormalization of
correlation functions of gauge-invariant local fields at
positive flow times can be shown to require no more than
a multiplicative renormalization of
the time-dependent quark fields once
the parameters of QCD are renormalized as usual.

For illustration, two applications of the extended flow 
are worked out in this paper,
one being a new strategy for the
calculation of the axial-current renormalization constant
in lattice QCD
and the other a computation of the chiral condensate
essentially
through the evaluation of the expectation
value of the scalar quark density at positive flow time.
In both cases, the use of the flow proves to be
technically attractive.
The chiral condensate, for example, is easily obtained with
high precision, because no additive renormalization is required.

In the next two sections, the extension of the gradient
flow to the quark fields is discussed in the continuum theory.
Since the flow equations respect chiral symmetry,
the correlation functions of the time-dependent fields
satisfy simple chiral Ward identities.
As explained in sect.~4, these allow
the correlation functions to be related to the 
physics of the light pseudo-scalar mesons.
Then follows a more technical part of the paper
(sects.~5--7), where the flow is set up in the framework of 
lattice QCD. In particular, the associated field theory
in 4+1 dimensions is shown to admit a well-defined
local lattice regularization. 
The viability of the proposed applications of the 
flow in numerical lattice QCD is finally demonstrated 
through a sample calculation in 2+1 flavour QCD (sects.~8,9).

\section 2. The gradient flow in QCD

The theory considered in this paper is QCD with gauge group $\SUn$ and a 
multiplet of two or more massive quarks in the 
fundamental representation of the gauge group. 
Many results are however of a fairly general nature
and not limited to QCD. The theory is set up in Euclidean
space and quantized through the functional integral as usual.
In this and the following section, dimensional regularization 
is employed.
The notational conventions are summarized
in appendix A.

\subsection 2.1 Flow equations

Let $A_{\mu}(x)$ be the $\SUn$ gauge potential,
$\psi(x)$ the quark field and $\psibar(x)$ the antiquark field
integrated over in the functional integral. The latter 
carry Dirac, colour and flavour indices, which are usually suppressed 
for simplicity. 

The Yang--Mills gradient flow evolves the gauge field as a function 
of a parameter $t\geq0$ that is referred to as the flow time.
Starting from the fundamental gauge field,
\equation{
  \left.B_{\mu}\right|_{t=0}=A_{\mu},
  \enum
}
the time-dependent field $B_{\mu}(t,x)$
is determined by the differential equation
\equation{
  \partial_tB_{\mu}=D_{\nu}G_{\nu\mu},
  \enum
  \nexteq{2ex}
  G_{\mu\nu}=\partial_{\mu}B_{\nu}-\partial_{\nu}B_{\mu}+[B_{\mu},B_{\nu}],
  \qquad
  D_{\mu}=\partial_{\mu}+[B_{\mu},\,\cdot\;]
  \enum
}
(see ref.~[\ref{WilsonFlow}] for an introduction to the subject).

As already mentioned in sect.~1, the extension of the flow to the
quark fields con\-sidered in this paper is a minimal one, where
the evolution of the gauge field is left unchanged.
The gauge field however appears in the flow 
equations\kern1pt\footnote{$\dagger$}{\footnotefont%
The left-action of any differential operator $\Delta$
(or of a difference operator in lattice field theory) 
is defined through the requirement that the relation
$\eta^{\dagger}\lvec{\Delta}=(\Delta\eta)^{\dagger}$
holds for all complex-valued fields $\eta$.}
\equation{
  \partial_t\chi=\Delta\chi,
  \qquad
  \partial_t\chibar=\chibar\lvec{\Delta},
  \enum
  \nexteq{2.5ex}
  \Delta=D_{\mu}D_{\mu},\qquad D_{\mu}=\partial_{\mu}+B_{\mu},
  \enum
}
which, together with the initial conditions
\equation{
  \left.\chi\right|_{t=0}=\psi,
  \qquad
  \left.\chibar\right|_{t=0}=\psibar,
  \enum
}
define the time-dependent quark and antiquark 
fields $\chi(t,x)$ and $\chibar(t,x)$.
In the flow equations (2.4), 
the gauge-covariant Laplacian $\Delta$ 
could be replaced by the square 
of the Dirac operator $\slash{D}$, for example, 
or be multiplied by a proportionality factor,
but such modifications
do not appear to offer any advantages.

The flow of quark fields introduced in this section 
is similar to the source 
smoothing operations proposed many years ago in
refs.~[\ref{Guesken},\ref{AlexandrouEtAl}].
With respect to these popular methods, there are, however, 
a few important differences,
one of them being the fact that the flow
operates in four dimensions rather than on the fields on an equal-time
hyper-plane. Moreover, the flow time varies continuously and the gauge field
is not set to the fundamental gauge field, but is
evolved together with the quark field.

\subsection 2.2 Local composite fields

Since the flow equations are gauge-covariant, the time-dependent
fields transform in the same way under gauge transformations as
the fundamental fields. Examples of 
gauge-invariant composite fields
that are local both in space-time and in flow time are the densities
\equation{
  E_t(x)=-\frac{1}{2}\tr\{G_{\mu\nu}(t,x)G_{\mu\nu}(t,x)\},
  \enum
  \nexteq{2.5ex}
  S^{rs}_t(x)=\chibar_r(t,x)\chi_s(t,x),
  \qquad
  P^{rs}_t(x)=\chibar_r(t,x)\dirac{5}\chi_s(t,x),
  \enum
}
where $r,s$ are flavour indices.

Through the initial conditions (2.1),(2.6), 
the time-dependent fields depend on the 
fundamental fields. 
From the point of view of the QCD functional 
integral, composite fields like the densities (2.7),(2.8)
are therefore observables, similar to Wilson loops or the 
ordinary pseudo-scalar densities, for example.
In the following, the quantities of
interest are the correlation functions of 
these fields,
i.e.~the expectation values of products of local fields composed 
from the basic fields at any flow time.

\subsection 2.3 Perturbation theory

As explained in refs.~[\ref{WilsonFlow},\ref{RenFlow}],
such correlation functions can be expanded in powers of the 
gauge coupling in a straightforward manner. 
First the flow equations are solved in powers of 
the fundamental fields. The substitution of 
these expansions in the local fields then leads to linear 
combinations
of correlation functions of the fundamental fields that
can be computed using the standard QCD Feynman rules.

In perturbation theory,
the flow equations (2.2) and (2.4) are replaced by
\equation{
  \partial_tB_{\mu}=D_{\nu}G_{\nu\mu}+\alpha_0D_{\mu}\partial_{\nu}B_{\nu},
  \enum
  \nexteq{2.5ex}
  \partial_t\chi=\Delta\chi-
  \alpha_0\partial_{\nu}B_{\nu}\chi,
  \qquad
  \partial_t\chibar=\chibar\lvec{\Delta}+
  \alpha_0\chibar\partial_{\nu}B_{\nu},
  \enum
}
in order to avoid some technical subtleties.
The terms proportional to the parameter $\alpha_0>0$ 
serve to damp the gauge modes
and could be removed by a time-dependent gauge transformation
[\ref{WilsonFlow}].
Correlation functions of gauge-invariant 
fields are therefore not affected by these extra terms.

The expansion of the time-dependent quark field $\chi(t,x)$ 
in powers of
the fundamental fields is obtained by iterating the 
quark flow equation in its integral form,
\equation{
  \chi(t,x)=\int\rmd^{D}\kern-1pt y\left\{K_t(x-y)\psi(y)
  +\int_0^t\rmd s\,K_{t-s}(x-y)\Delta'\chi(s,y)\right\},
  \enum
  \nexteq{3.0ex}
  \Delta'=(1-\alpha_0)\partial_{\nu}B_{\nu}
  +2B_{\nu}\partial_{\nu}+B_{\nu}B_{\nu},
  \enum
}
and the corresponding integral
equation for the time-dependent gauge field
[\ref{WilsonFlow}], where
\equation{
  K_t(z)={\rme^{-z^2/4t}\over(4\pi t)^{D/2}}
  \enum
}
denotes the heat kernel of the Laplacian in $D$ dimensions. 
At the lowest order in the gauge coupling, 
the interaction term in eq.~(2.11) can be dropped
and the expression
\equation{
  \langle\chi(t,x)\chibar(s,y)\rangle=
  \int{\rmd^{D}\kern-1pt p\over(2\pi)^D}\,
  \rme^{ip(x-y)}
  {\rme^{-(t+s)p^2}\over M_0+i\slash{p}}+\rmO(g_0^2)
  \enum
}
is then obtained for the two-point function of the time-dependent
quark field, where $g_0$ and $M_0$ are the bare coupling 
and (diagonal) quark mass matrix.
The smoothing character of the quark flow
is evident from these equations.
Moreover,
the smoothing range at leading order, $\sqrt{8t}$, 
is seen to be the same as in the case of the gauge field.

There are $8$ self-energy diagrams that contribute
to the two-point function (2.14) at one-loop order of perturbation theory.
The vertices in these diagrams derive from the QCD action
and from the iteration of the integral equation (2.11).
Following the lines of ref.~[\ref{RenFlow}], it is straightforward
to compute the ultraviolet divergent parts of the diagrams and
one then finds that the two-point function can be renormalized
by renormalizing the quark masses as usual and 
the fields according to
\equation{
  \chi=Z_{\chi}^{-1/2}\ren{\chi},
  \qquad
  \chibar=\ren{\chibar}Z_{\chi}^{-1/2}.
  \enum
}
In the $\MSbar$ scheme in $D=4-2\eps$ dimensions,
the calculation yields
\equation{
  Z_{\chi}=1+{3\CF\over16\pi^2\eps}g^2+\rmO(g^4),
  \qquad
  \CF={N^2-1\over2N},
  \enum
}
for the field renormalization constant,
$g$ being the renormalized coupling.

\subsection 2.4 Quark condensate at non-zero flow time

At non-zero flow times, the large momenta in the integral (2.14) 
are exponentially suppressed and the quark two-point function 
consequently has no singularities as
$(t,x)\to(s,y)$. The absence of short-distance singularities
is a general feature of the correlation functions at positive
flow times, which derives from the smoothing property of
the flow equations. 
In particular, the ``time-dependent quark condensates''
\equation{
  \Sigt^{rr}=-\langle S^{rr}_t(x)\rangle
  \enum
}
do not require additive renormalization.

In perturbation theory,
the renormalized condensates 
\equation{
  \rens{\Sigma}{t}^{rr}=Z_{\chi}\Sigt^{rr}
  \enum
}
can be worked out in powers of the renormalized coupling $g$,
with coefficients that depend on the renormalized quark masses 
$\mR{r},\mR{s},\ldots$ and the flow time $t$.
The leading-order term in four dimensions is given by
\equation{
  \left.\rens{\Sigma}{t}^{rr}\right|_{g=0}=
  {2N\mR{r}\over(4\pi)^2t}
  \int_0^{\infty}\rmd v\,{\rme^{-vz}\over(1+v)^2},
  \qquad z=2t\mR{r}^2,
  \enum
}
and thus vanishes in the chiral limit,
consistently with the fact that chiral symmetry can only
be spontaneously broken at the non-perturbative level.

\section 3. Field theory in $\mib D$+1 dimensions

The discussion in the following sections heavily builds 
on the fact that the correlation functions 
of the time-dependent fields coincide with 
the correlation functions in
a local field theory in $D$+1 dimensions,
the extra dimension being the flow time.
The observation dates back to the seminal work of
Zinn--Justin and Zwanziger [\ref{Zinn},\ref{ZinnZwanziger}]
on the Langevin equation and recently
allowed the renormalizability 
of the (pure-gauge) gradient flow to be established 
to all orders of perturbation theory
[\ref{RenFlow}]. 

As will become clear below, the inclusion of the quark fields 
in the theory in $D$+1 dimensions is straightforward.
The general setup is exactly the same as in ref.~[\ref{RenFlow}],
where an introduction to the subject and further details 
can be found.

\subsection 3.1 Fields and action

In addition to the fundamental and the time-dependent fields
already encountered, the theory in $D$+1 dimensions
involves the Lagrange-multiplier fields $L_{\mu}(t,x)$, $\lambda(t,x)$
and $\lambdabar(t,x)$. The latter are fermion fields with
the same indices as the quark fields, while $iL_{\mu}(t,x)$ is a
vector field that takes values in the Lie algebra of $\SUn$. 

As explained in ref.~[\ref{RenFlow}], 
gauge fixing in perturbation theory requires the introduction of 
the Faddeev--Popov ghost
fields $c(x),\cbar(x)$ and the associated 
time-dependent fields $d(t,x),\dbar(t,x)$,
where
\equation{
  \left.d\right|_{t=0}=c.
  \enum
}
No boundary condition is imposed on the field $\dbar(t,x)$, 
which, in many ways, plays a 
r\^ole similar to the Lagrange-multiplier fields.
Apart from the boundary conditions (2.1),(2.6) and (3.1),
all fields are
assumed to be unconstrained at this point.

The total action of the theory,
\equation{
  S_{\rm tot}=S+\SGfl+\SFPfl+\SFfl,
  \enum
}
includes the gauge-fixed QCD action $S$ of the fields at 
flow time zero and the bulk actions
\equation{
  \SGfl=-2\int_0^{\infty}\rmd t\int\rmd^D\kern-1pt x\, 
  \tr\bigl\{L_{\mu}(t,x)
  \bigl(\partial_tB_{\mu}-
        D_{\nu}G_{\nu\mu}-\alpha_0D_{\mu}\partial_{\nu}B_{\nu}
  \bigr)(t,x)\bigr\},
  \enum
  \nexteq{3.0ex}
  \SFPfl=-2\int_{0}^{\infty}\rmd t\int\rmd^D\kern-1pt x\,
  \tr\bigl\{
  \dbar(t,x)\bigl(\partial_td-\alpha_0 D_{\mu}\partial_{\mu}d\bigr)(t,x)
  \bigr\},
  \enum
  \nexteq{3.0ex}    
  \SFfl=\int_0^{\infty}\rmd t\int\rmd^D\kern-1pt x\,
  \bigl\{\lambdabar(t,x)
         (\partial_t-\Delta+\alpha_0\partial_{\nu}B_{\nu})\chi(t,x)
  \noenum
  \nexteq{1.5ex}
  {\phantom{\SFfl=\int_0^{\infty}\rmd t\int\rmd^D\kern-1pt x\,\bigl\{}}
  +\chibar(t,x)
  (\lvec{\partial}_t-\lvec{\Delta}-\alpha_0\partial_{\nu}B_{\nu})
  \lambda(t,x)\bigr\}.
  \enum
}
With respect to the pure-gauge flow studied
in ref.~[\ref{RenFlow}], 
the bulk quark action $\SFfl$ is the only new term in the 
total action.

\subsection 3.2 Correlation functions

The $n$-point correlation functions of the 
gauge, quark, ghost and Lagrange-multiplier fields 
are formally defined through the functional integral
over all fields with ``Boltzmann factor'' 
$\exp\{-S_{\rm tot}\}$. 

In the case of the correlation functions of the gauge and quark fields,
the functional integral can be worked out, to some
extent, by performing the integral over the Lagrange-multiplier
fields. Since the action is linear in these fields,
the integration yields 
a product of functional Dirac $\delta$-functions.
The bulk actions were chosen such that
the $\delta$-functions amount to imposing the flow equations on 
the time-dependent gauge and quark fields. 
Up to mathematical subtleties and possibly non-trivial Jacobian factors,
the correlation functions thus coincide with the ones
calculated directly, as in sect.~2, by
solving the flow equations with the initial
conditions (2.1),(2.6) and treating the calculated
time-dependent fields as QCD observables.

The equivalence of the theory in $D$+1 dimensions and the 
direct computation of the correlation functions
can be rigorously shown in perturbation theory
[\ref{RenFlow}] and in lattice QCD (sect.~5).
In particular, all correlation functions 
that do not involve the time-dependent quark fields or
the associated Lagrange-multiplier fields
coincide with the ones studied in ref.~[\ref{RenFlow}],
where the quark fields were not evolved in flow time.
This ``reduction property'' 
of the theory in $D$+1 dimensions strongly constrains
the form of the possible counterterms
required for renormalization (see subsect.~3.4).

\subsection 3.3 Fermion integral

Since the action in $D$+1 dimensions 
is quadratic in the fermion fields,
the functional integral over these fields can be
evaluated (at fixed bosonic fields) by applying Wick's theorem.
The basic Wick contractions are worth being given explicitly, 
because the expressions provide some insight
into the structure of the theory
in $D$+1 dimensions and will, in any case, be
frequently needed later.

The Wick contractions of the basic fields
are determined by the action and the boundary conditions (2.6).
In particular, the propagator of
the quark fields at flow time zero,
\equation{
  \wick{\psi(x)\psibar(y)}=S(x,y),
  \enum
  \nexteq{2.5ex}
  (\slash{D}+M_0)S(x,y)=\delta(x-y),
  \enum
}
coincides with the one in $D$ dimensions.
Apart from the quark propagator $S(x,y)$, 
the contractions of the time-dependent fermion fields,
\equation{
  \longwick{\lambda(t,x)\lambdabar(s,y)}=0,
  \enum
  \nexteq{2.5ex}
  \longwick{\chi(t,x)\lambdabar(s,y)}=\theta(t-s)K(t,x;s,y),
  \enum
  \nexteq{2.5ex}
  \longwick{\lambda(t,x)\chibar(s,y)}=\theta(s-t)K(s,y;t,x)^{\dagger},
  \enum
  \nexteq{2.5ex}
  \longwick{\chi(t,x)\chibar(s,y)}=\int\rmd^Dv\,
  \rmd^Dw\,
  K(t,x;0,v)S(v,w)K(s,y;0,w)^{\dagger},
  \enum
}
involve the fundamental solution
\equation{
  \{\partial_t-\Delta+
  \alpha_0\partial_{\nu}B_{\nu}\}K(t,x;s,y)=0\quad\hbox{if}\quad t\geq s,
  \enum
  \nexteq{2.0ex}
  \lim_{t\to s}K(t,x;s,y)=\delta(x-y),
  \enum
}
of the quark flow equation. Note that the
kernel $K(t,x;s,y)$ is flavour-independent, proportional
to the unit Dirac matrix and 
an $N\times N$ complex matrix in colour space (the adjoint of
the kernel in eqs.~(3.10) and (3.11) is 
taken in index space only). 

Consistently with the boundary conditions at vanishing flow time,
the contractions 
involving the fundamental quark and antiquark fields,
\equation{
  \longwick{\chi(t,x)\psibar(y)}=\int\rmd^Dv\,K(t,x;0,v)
  S(v,y),
  \enum
  \nexteq{2.5ex}
  \wick{\psi(x)\chibar(s,y)}=\int\rmd^Dw\,S(x,w)
  K(s,y;0,w)^{\dagger},
  \enum
  \nexteq{2.5ex}
  \longwick{\lambda(t,x)\psibar(y)}=
  \wick{\psi(x)\lambdabar(s,y)}=0,
  \enum
}
are included in eqs.~(3.9)--(3.11) as special cases.
Note, however, that the behaviour of 
correlation functions 
near space-time boundaries tends to 
be complicated by short-distance singularities
[\ref{Symanzik}]. 
The limits
\equation{
  \lim_{s\to0}\longwick{\lambda(0,x)\chibar(s,y)}=
  \lim_{t\to0}\longwick{\chi(t,x)\lambdabar(0,y)}=
  \delta(x-y),
  \enum
}
for example,
differ from the contractions (3.16) at $t=s=0$ by a contact term.

\subsection 3.4 Renormalization

In perturbation theory, the renormalization properties of
the theory in $D$+1 dimensions can
be determined using power counting and 
symmetry arguments.
The task of figuring out all possible local counterterms
to the action is simplified by the fact that the renormalization of
the theory where only the gauge field is evolved in flow time
is already known to require no more than the usual
QCD parameter and field renormalizations [\ref{RenFlow}]. In particular,
at flow time $t>0$,
the fields $B_{\mu}(t,x)$ and $L_{\mu}(t,x)$ need not
be renormalized. 

When the time-dependent quark fields are included in the theory,
further counter\-terms may need to be added to the total action.
The reduction property mentioned at the end of subsect.~3.2
however implies that any such term
must be proportional to the quark fields or
the associated Lagrange-multiplier fields.
Moreover,
as explained in subsect.~7.2 of ref.~[\ref{RenFlow}], 
bulk counterterms (i.e.~local terms integrated over all
$D$+1 coordinates)
are excluded, because
the correlation functions at large flow times
are given by tree diagrams. 

Taking the symmetries of the unrenormalized theory 
into account, and the fact
that the quark and Lagrange-multiplier fields have
power-counting dimension $3/2$ and $5/2$, respectively,
a moment of thought reveals that the only possible additional
counterterm is proportional to
\equation{
  \int\rmd^D\kern-1pt x\,
  \left\{\lambdabar(0,x)\psi(x)+\psibar(x)\lambda(0,x)\right\}.
  \enum
}  
The insertion of this term in the Feynman diagrams
is equivalent to a renormalization 
\equation{
  \chi=Z_{\chi}^{-1/2}\ren{\chi},
  \qquad
  \lambda=Z_{\chi}^{1/2}\ren{\lambda},
  \enum
}
of the fermion and, similarly, the anti-fermion fields
at positive flow time.
Apart from the usual renormalization of the 
QCD parameters and fields,
the renormalization of the theory in $D$+1
dimensions thus requires the time-dependent quark
and associated Lagrange multiplier fields 
to be renormalized as specified above.

\section 4. Chiral symmetry relations

Since the quark flow equations (2.4) preserve chiral symmetry,
the time-dependent composite fields like the densities (2.8) 
transform in a simple way under chiral rotations
and thus allow the spontaneous breaking of the symmetry
to be studied in useful new ways.
In this section, the principal goal is to relate
the time-dependent condensates (2.17) to the 
ordinary chiral condensate and the pseudo-scalar decay
constant in the chiral limit.

\subsection 4.1 Quark field equation and PCAC relation

The space-time dimension is now set to 4
and the need for regularization and renormalization
is ignored for the moment, i.e.~the statements made 
in this subsection
are, strictly speaking, only valid
at tree level of perturbation theory. 

Perhaps somewhat unexpectedly, the quark field equation
in the theory in 4+1 dimensions,
\equation{
  \langle\{(\slash{D}+M_0)\psi(x)-\lambda(0,x)\}\kern1pt
  \phi_1(t_1,x_1)\ldots\phi_n(t_n,x_n)
  \rangle=
  \hbox{contact terms},
  \enum
}
differs from the one
in QCD by a term that couples to the quark fields at non-zero
flow time. Equation (4.1) holds for any
local fields $\phi_1,\ldots,\phi_n$ and the contact terms vanish
if all points $(t_1,x_1),\ldots,(t_n,x_n)$ in 4+1 dimensions 
are different from $(0,x)$.
The validity of the field equation (4.1) follows
from the identities
\equation{
  (\slash{D}+M_0)\wick{\psi(x)\psibar(y)}
  =\longwick{\lambda(0,x)\psibar(y)}+\delta(x-y),
  \enum
  \nexteq{2.5ex}
  (\slash{D}+M_0)\wick{\psi(x)\chibar(s,y)}
  =\longwick{\lambda(0,x)\chibar(s,y)},
  \enum
  \nexteq{2.5ex}
  (\slash{D}+M_0)\wick{\psi(x)\lambdabar(s,y)}
  =\longwick{\lambda(0,x)\lambdabar(s,y)},
  \enum
}
which are a direct consequence of the formulae quoted
in subsect.~3.3 for the Wick contractions of the fermion fields.
Equations (4.2)--(4.4) moreover show that contact terms only arise 
from contractions of the fundamental quark fields.

Now let 
\equation{
  \Aax^{rs}_{\mu}(x)=\psibar_r(x)\dirac{\mu}\dirac{5}\psi_s(x),
  \qquad
  \Pax^{rs}(x)=\psibar_r(x)\dirac{5}\psi_s(x),
  \enum
}
be the axial currents and densities in QCD.
For $r\neq s$, the quark field equation then implies the
PCAC relation
\equation{
  \langle
  \{\partial_{\mu}\Aax^{rs}_{\mu}(x)-(m_{0,r}+m_{0,s})\Pax^{rs}(x)
  +\Ptax^{rs}(x)\}\kern1pt
  \phi_1(t_1,x_1)\ldots\phi_n(t_n,x_n)\rangle
  \noenum
  \nexteq{2.0ex}
  \quad=\hbox{contact terms},
  \enum
}
where $m_{0,r},m_{0,s}$ are the bare masses of the quarks 
with flavour labels $r,s$ and
\equation{
  \Ptax^{rs}(x)=\lambdabar_r(0,x)\dirac{5}\psi_s(x)
  +\psibar_r(x)\dirac{5}\lambda_s(0,x).
  \enum
}
In particular, the identity
\equation{
  \langle
  \{\partial_{\mu}\Aax^{rs}_{\mu}(x)-(m_{0,r}+m_{0,s})\Pax^{rs}(x)
  +\Ptax^{rs}(x)\}\kern1pt P^{sr}_t(y)\rangle=0
  \enum
}
holds at all flow times $t>0$ and all points $x$ including $x=y$.
There are no contact terms in this case, because
the fermion propagators that contribute to the 
correlation function on the left of eq.~(4.8) all go
from $(0,x)$ to $(t,y)$ in 4+1 dimensions or from $(t,y)$ to $(0,x)$.
The contraction (4.2) thus does not occur. 
In view of 
the smoothing character of the flow equations, all parts
of the correlation function are in fact regular functions 
of $x$.

Another purely algebraic consequence of the structure of the 
Wick contractions is the relation
\equation{
  \int\rmd^4x\,\langle\Ptax^{rs}(x)P^{sr}_t(y)\rangle
  =\langle S^{rr}_t(y)\rangle+\langle S^{ss}_t(y)\rangle,
  \enum
}
which, when combined with the PCAC relation (4.8),
leads to an identity
\equation{
  (m_{0,r}+m_{0,s})\int\rmd^4x\,\langle\Pax^{rs}(x)P^{sr}_t(y)\rangle
  =-\Sigt^{rr}-\Sigt^{ss}
  \enum
}
that allows the time-dependent quark condensates to be 
related to the physics of the pseudo-scalar mesons
(see subsect.~4.3). In the derivation of this equation,
both the quark masses and the flow time 
were assumed to be positive, as otherwise
the absence of boundary terms at large $x$
and non-integrable singularities at $x=y$ 
would not be guaranteed.

\subsection 4.2 Renormalized PCAC relation

Beyond tree level of perturbation theory, the fields and quark masses
in eqs.~(4.8)--(4.10) require renormalization. The equations then become
relations among renormalized correlation functions, whose exact
form depends on the chosen normalization conventions.
There are some natural choices one can make here
and it is the goal in the following lines to specify these.
The discussion implicitly assumes a sensible regularization
of the theory, such as dimensional regularization or 
Wilson's lattice formulation [\ref{Wilson}],
where most symmetries are preserved. As before, only the 
flavour non-diagonal channels are considered.

The
non-singlet axial currents and densities
renormalize multiplicatively,
\equation{
  \rens{\Aax}{\mu}^{rs}=Z_A\Aax_{\mu}^{rs},
  \qquad
  \ren{\Pax}^{rs}=Z_P\Pax^{rs},
  \enum
}
and the same is the case for the time-dependent densities
\equation{
  \rens{S}{t}^{rs}=Z_{\chi}S_t^{rs},
  \qquad
  \rens{P}{t}^{rs}=Z_{\chi}P_t^{rs}.
  \enum
}
As already mentioned, since there are
no short-distance singularities at positive flow time,
these fields (including the flavour diagonal
ones) renormalize according to their quark content, i.e.~like 
the product of a time-dependent quark and antiquark field at non-zero 
distance in 4+1 dimensions.

In the case of the field $\Ptax^{rs}$,
the multiplicative renormalizability,
\equation{
  \ren{\Ptax}^{rs}=\tilde{Z}_P\Ptax^{rs},
  \enum
}
is less obvious,
because the field has dimension $4$ and could mix
with other fields of dimension $3$ and $4$.
The Wick contractions involving the
Lagrange-multiplier fields $\lambda$ and $\lambdabar$ 
are however such that, up to contact terms,
the two-point correlation functions of
$\Ptax^{rs}$ and any local 
field composed from the gauge and quark fields 
at flow time zero vanish. 
Divergent additive renormalizations by such fields are therefore
excluded. One is then left with
fields involving the Lagrange-multiplier fields, but 
since $\Ptax^{rs}$ is the only one among these with
dimension 4 and the required symmetry properties,
additive renormalizations are again not possible
and the field must consequently renormalize multiplicatively.

The normalizations of the renormalized fields and 
the renormalized quark masses $\mR{r}$ may now be chosen such that
the PCAC relation assumes the form
\equation{
  \langle
  \{\partial_{\mu}\rens{\Aax}{\mu}^{rs}(x)-
  (\mR{r}+\mR{s})\ren{\Pax}^{rs}(x)
  +\ren{\Ptax}^{rs}(x)\}\kern1pt\rens{P}{t}^{sr}(y)\rangle=0.
  \enum
}  
Note that this convention only fixes the relative normalization of 
the axial current $\rens{\Aax}{\mu}^{rs}$, the product
$(\mR{r}+\mR{s})\ren{\Pax}^{rs}$ and the density $\ren{\Ptax}^{rs}$.
The normalization of the latter may however
be fixed by requiring the equation
\equation{
  \int\rmd^4x\,\langle\ren{\Ptax}^{rs}(x)\rens{P}{t}^{sr}(y)\rangle
  =\langle\rens{S}{t}^{rr}(y)\rangle+\langle\rens{S}{t}^{ss}(y)\rangle
  \enum
}
and thus the identity
\equation{
  (\mR{r}+\mR{s})\int\rmd^4x\,
  \langle\ren{\Pax}^{rs}(x)\rens{P}{t}^{sr}(y)\rangle
  =-\rens{\Sigma}{t}^{rr}-\rens{\Sigma}{t}^{ss}
  \enum
}
to hold (where $\rens{\Sigma}{t}^{rr}=Z_{\chi}\Sigma_t^{rr}$).
Further normalization conditions then still need to be imposed
on the renormalized 
quark masses and the time-dependent densities, but for the masses
one can adopt one of the standard conventions, while the normalization of
the time-dependent fields tends to cancel in the equations of interest 
and is left unspecified.

The normalization of the axial currents
implicitly determined by eqs.~(4.14),(4.15) coincides with
the one usually chosen, where the axial charges
assume integer values. There are different ways to 
show this, the perhaps most straightforward one starting from
a formulation of lattice QCD that preserves chiral symmetry 
[\ref{GinspargWilson}--\ref{ExactChSy}].
Using chiral Ward identities [\ref{BoulderReview}],
the bare fields can be shown to satisfy eqs.~(4.8),(4.9) in this case,
up to terms vanishing proportionally to the lattice spacing.
Equations (4.14),(4.15) then imply
$Z_A=\tilde{Z}_P=1$ and thus the canonical normalization of 
the renormalized axial currents. In view of the universality of the
continuum limit,
the canonical normalization is therefore guaranteed 
by these equations independently 
of the chosen regularization of the theory.

\subsection 4.3 Pseudo-scalar matrix elements and the chiral condensate

The quark multiplet is now assumed to contain two light quarks, 
referred to as the $u$ and $d$ quarks, with the same
renormalized mass $\ren{m}$.
If there are not too many further
quarks, chiral symmetry in the $(u,d)$-channel 
is expected to be spontaneously broken and there is a triplet of 
pseudo-scalar mesons, the ``pions", whose mass vanishes
as $\ren{m}$ goes to zero.

At large distances, the light-quark pseudo-scalar correlation
functions are dominated by the intermediate one-pion states.
In particular, as $x_0\to\infty$
\equation{
  \int\rmd^3{\mib x}\,\langle\ren{\Pax}^{ud}(x)\ren{\Pax}^{du}(0)\rangle
  =-{\Gpi^2\over\Mpi}
  \rme^{-\Mpi x_0}\bigl\{1+\rmO(\rme^{-\Delta Ex_0})\bigr\},
  \enum
  \nexteq{2.5ex}
  \int\rmd^3{\mib x}\,\langle\rens{\Aax}{0}^{ud}(x)\ren{\Pax}^{du}(0)\rangle
  =\Fpi\Gpi\rme^{-\Mpi x_0}\bigl\{1+\rmO(\rme^{-\Delta Ex_0})\bigr\},
  \enum
}
where $\Delta E$ denotes the energy gap to the higher states and
$\Mpi$, $\Fpi$ and $\Gpi$ are the pion mass, 
decay constant and vacuum-to-pion matrix element of the axial 
density.

From the point of view of the theory in 4 dimensions,
$\rens{\Pax}{t}^{du}(x)$ is not a local field, but
it is still a well localized expression in the fundamental fields
since the smoothing kernel $K(t,x;0,y)$ falls off approximately 
like a Gaussian at large distances $|x-y|$.
At times $x_0$ much larger than the smoothing range $\sqrt{8t}$, 
the two-point function
\equation{
  \int\rmd^3{\mib x}\,\langle\ren{\Pax}^{ud}(x)\rens{\Pax}{t}^{du}(0)\rangle
  =-{\Gpi\Gpit\over\Mpi}
  \rme^{-\Mpi x_0}\bigl\{1+\rmO(\rme^{-\Delta Ex_0})\bigr\}
  \enum
}
therefore decays in the same way as the ordinary pseudo-scalar
correlation functions.
The coefficient $\Gpit$ coincides with
the vacuum-to-pion matrix element of a certain operator,
but is here considered to be defined through eq.~(4.19).

The low-energy effective constants characterizing the 
(two-flavour) chiral limit,
\equation{
  \Sigma=\lim_{m_{\rm R}\to0}\Fpi\Gpi,
  \enum
  \nexteq{2.5ex}
  F=\lim_{m_{\rm R}\to0}\Fpi,
  \enum
}
are the chiral condensate $\Sigma$
and the value $F$ of the pion decay constant at $\ren{m}=0$.
As already mentioned, these can be related to 
the time-dependent light-quark con\-den\-sate
$\rens{\Sigma}{t}=\rens{\Sigma}{t}^{uu}$ via
the chiral Ward identity (4.16). Using the PCAC
relation
\equation{
  2\ren{m}\Gpi=\Mpi^2\Fpi,
  \enum
}
the latter may be written in the form
\equation{
  \rens{\Sigma}{t}=-{\Mpi^2\Fpi\over2\Gpi}
  \int\rmd^4x\,\langle \ren{P}^{ud}(x)\rens{P}{t}^{du}(0)\rangle.
  \enum
}
The integral on the right of this equation 
diverges when $\ren{m}\to0$ and its asymptotic behaviour is determined
by the pion pole of the correlation function in momentum space.
Recalling eq.~(4.19), one is then led to conclude that
\equation{
  \Sigma=\lim_{m_{\rm R}\to0}{\rens{\Sigma}{t}\Gpi\over\Gpit},
  \enum
  \nexteq{2.5ex}
  F=\lim_{m_{\rm R}\to0}{\rens{\Sigma}{t}\over\Gpit}.
  \enum
}
In the chiral limit, the time-dependent condensate $\rens{\Sigma}{t}$
thus provides an order parameter for the spontaneous
breaking of chiral symmetry, which coincides with the standard
condensate $\Sigma$ up to a
proportionality constant.

A remarkable feature of eqs.~(4.24),(4.25) is the fact that
there is no reference to the axial currents, not even implicitly 
via the 
normalization conditions, and that the equations hold for 
any (positive) flow time $t$.
Moreover, the renormalization constant 
$Z_{\chi}$ drops out so that the decay constant 
$F$, for example, is directly given through the bare time-dependent 
condensate and pseudo-scalar vacuum-to-pion matrix element.

\subsection 4.4 Chiral perturbation theory

Close to the chiral limit, the asymptotic behaviour of 
many quantities in QCD can be described by chiral perturbation theory.
If the Compton wavelength of the pion is much larger than
the smoothing range of the gradient flow, i.e.~if
\equation{
  8t\Mpi^2\ll1,
  \enum
}
the time-dependent densities $\rens{S}{t}^{rs}$ and $\rens{P}{t}^{rs}$
are local fields from the point of view of pion physics, with
the same symmetry properties as the densities at vanishing flow time.
At low energies, the correlation functions of the time-dependent 
fields and the ordinary axial currents and densities are
therefore expected to be dominated by the 
pion intermediate states and should hence be accessible to a quantitative
description in the framework of chiral perturbation theory.

The time-dependent densities can in fact easily be included in 
chiral perturbation theory by adding the appropriate source terms 
to the chiral Lagrangian [\ref{ChPT}]. At each order of the 
chiral expansion, these terms require a number of effective couplings
to be introduced, which will, in general, depend on the flow time.
The expansion of the chiral condensate $\rens{\Sigma}{t}$
and the matrix element $\Gpit$, for example, 
can then be worked out
and provides insight into how exactly the
chiral limits (4.24),(4.25) are reached.

\section 5. Lattice regularization

The starting point in this section is a formulation of lattice
QCD, where the quark fields are put on the lattice as proposed
by Wilson [\ref{Wilson}]. 
No particular assumptions
are made on the lattice Dirac operator or the gauge action,
but both should be local and respect all symmetries preserved
by the standard Wilson theory. For notational convenience,
the lattice spacing $a$ is set to unity.
The lattice is taken to be infinitely extended in all directions 
unless specified otherwise.

\subsection 5.1 Flow equations

On the lattice, the time-dependent gauge field $V(t,x,\mu)$ is
defined by 
\equation{
  \left.V\right|_{t=0}=U,
  \enum
  \nexteq{2.5ex}
  \partial_tVV^{-1}=-g_0^2\partial\Sw,
  \enum
}
where $U(x,\mu)$ is the fundamental lattice gauge field 
and $(\partial\Sw)(V,x,\mu)$ the gradient of the
Wilson plaquette action (see appendix A and
ref.~[\ref{WilsonFlow}] for further explanations).

In the case of the quark fields, the boundary conditions 
and the form of the flow equations are the same as 
in the continuum theory, but the operator $\Delta$
in eq.~(2.4) gets replaced by the lattice Laplacian
\equation{
  \Delta=\nabstar{\mu}\nab{\mu},
  \enum
}
$\nab{\mu}$ and $\nabstar{\mu}$
being the gauge-covariant forward and backward
difference operators in presence of the time-dependent gauge field.

As already noted in ref.~[\ref{WilsonFlow}],
the global existence, uniqueness and smoothness of the solution
of the flow equation (5.2) is rigorously guaranteed.
Since $\Delta$ is a bounded operator, and since
it depends smoothly on the flow time, the 
same can be shown to be true in the case of quark flow equations
[\ref{ArnoldI}].
Local fields at non-zero flow time, such as those considered
in the previous sections, are thus completely well defined
on the lattice.

\subsection 5.2 Lattice theory in 4+1 dimensions

Similarly to the continuum theory,
the lattice QCD correlation functions of the 
time-dependent local fields can be obtained
from a lattice field theory in 4+1 dimensions.
To be able to fully control this construction,
the latter is first set up for a discretized
version of the flow equations, where the flow time is restricted
to a finite set
\equation{
  0,\eps,2\eps,3\eps,\ldots,T
  \enum
}
of values
separated by a time step $\eps>0$. 

The integration of the discretized evolution equation 
\equation{
  V(t+\eps,x,\mu)V(t,x,\mu)^{-1}=\rme^{-\eps g_0^2(\partial\Sw)(V,x,\mu)}
  \enum
}
for the gauge field amounts to an approximate integration of the 
continuous equation (5.2) using the Euler method
(or, equivalently, through the application
of a sequence of ``stout smearing'' steps [\ref{Stout}]).
In the case
of the quark fields, the discretized flow equation has the 
same form as the continuous one, but the time derivative 
of the quark field is replaced by the forward difference
\equation{
  \partial_t\chi(t,x)={1\over\eps}\{\chi(t+\eps,x)-\chi(t,x)\}.
  \enum
}
Note that the discrete equations are such that they can be
solved in steps,
starting from the fundamental fields at time $t=0$
and recursively proceeding from time $t$ to $t+\eps$.
In particular, the solutions are uniquely determined by the
initial values of the fields.

In the lattice theory in 4+1 dimensions, all fields integrated
over in the functional integral are defined at the times (5.4)
only. 
Since gauge fixing is not needed in the present context,
there are no ghost fields and no gauge terms in the action,
but the boundary conditions (2.6) and (5.1) are imposed 
as before.
The components of the Lagrange-multiplier fields at time $T$ 
decouple from the other fields and can therefore be set
to zero (or be omitted from the beginning). 

A possible choice
of the bulk actions is then
\equation{
  \SGfl=-2\sum_{t=0}^{T-\eps}\sum_{x,\mu}
  \tr\bigl\{L(t,x,\mu)
  \noenum
  \nexteq{-0.5ex}
  {\phantom{\SGfl=-2\sum_{t=0}^{T-\eps}\sum_{x,\mu}}}
  \times
  \Psu\bigl(
  V(t+\eps,x,\mu)V(t,x,\mu)^{-1}-\rme^{-\eps g_0^2(\partial\Sw)(V,x,\mu)}
  \bigr)\bigr\},
  \enum
  \nexteq{2.5ex}
  \SFfl=\eps\sum_{t=0}^{T-\eps}\sum_x\bigl\{
  \lambdabar(t,x)\left(\partial_t-\Delta\right)\chi(t,x)
  +\chibar(t,x)\left(\lvec{\partial}_t-\lvec{\Delta}\right)\lambda(t,x)
  \bigr\},
  \enum
}
where the function
\equation{
  M\to\Psu(M)={1\over2}(M-M^{\dagger})
  -{1\over2N}\kern1pt\tr(M-M^{\dagger})
  \enum
}
maps any complex $N\times N$ matrix to an element of the Lie algebra
of $\SUn$. Apart from these actions and the lattice QCD action,
the measure term
\equation{
  \Sms=\sum_{t=0}^{T-\eps}\sum_{x,\mu}
  {\cal L}_{\rm ms}(V(t+\eps,x,\mu)V(t,x,\mu)^{-1}),
  \enum
  \nexteq{3.0ex}
  {\cal L}_{\rm ms}(W)=
  \cases{-\ln\det(J(W)) & if $W\in\Dsu$,\cr
         \noalign{\vskip1ex}
         \infty         & otherwise,\cr}
  \enum
}
must be included in the total action of the lattice theory (see appendix B
for the definition of the Jacobian matrix $J(W)^{ab}$ of the mapping (5.9) 
and the neighbourhood $\Dsu\subset\SUn$ of unity). 

Having specified the fields and their action,
the functional integral of the lattice theory in 4+1 dimensions
is defined in the standard manner. 
Note that the measure term (5.10),(5.11) 
effectively restricts the link variables 
in the functional integral to the domain where
\equation{
  V(t+\eps,x,\mu)V(t,x,\mu)^{-1}\in\Dsu
  \quad\hbox{for all $t,x,\mu$}.
  \enum
}
Since the action $\SGfl$ is purely imaginary, 
the integral over the Lagrange-multiplier field $L(t,x,\mu)$ 
is not absolutely convergent, but it is understood
that the integration ambiguity this may entail is removed 
through the addition of an
infinitesimal quadratic term in the Lagrange-multiplier field
to the action.

\subsection 5.3 Recovering the theory in four dimensions

The correlation functions defined in this way of local fields 
composed from the fields $V$, $\chi$ and $\chibar$
can be evaluated, to some extent,
by first integrating over the Lagrange-multiplier fields.
Since the total action is linear in the latter, 
the integration yields a product
\equation{
  \prod_{t=0}^{T-\eps}\prod_{x,\mu}
  \delta\bigl\{\Psu\bigl(
  V(t+\eps,x,\mu)V(t,x,\mu)^{-1}\bigr)-
  \Psu\bigl(\rme^{-\eps g_0^2(\partial\Sw)(V,x,\mu)}\bigr)\bigr\}
  \noenum
  \nexteq{1.0ex}
  \times\prod_{t=0}^{T-\eps}\prod_{x}
  \delta\bigl\{\partial_t\chi(t,x)-\Delta\chi(t,x)\bigr\}
  \delta\bigl\{\chibar(t,x)\lvec{\partial}_t-
  \chibar(t,x)\lvec{\Delta}\bigr\}
  \enum
}
of Dirac $\delta$-functions.
The arguments of these 
$\delta$-functions vanish if $V$, $\chi$ and $\chibar$ satisfy the discrete
flow equations, and for sufficiently small step sizes $\eps$
there is in fact no other way the arguments 
can be made to vanish%
\kern1pt\footnote{$\dagger$}{\footnotefont%
Uniqueness is guaranteed if 
$\exp\{\eps g_0^2(\partial\Sw)(V,x,\mu)\}\in\Dsu$
on all links of the lattice and at all flow times.
Some rigorous estimates show that
this condition is satisfied
if $\eps\leq(24\sqrt{2}N)^{-1}$.}.
In the case of the gauge field,
the important point to note here is
that the mapping $\Psu$ in the $\delta$-functions 
operates on $\SUn$ matrices in a small neighbourhood of unity.
The properties of the mapping quoted in appendix B then
imply that the constraints imposed by the $\delta$-functions 
are equivalent to imposing the flow equation (5.5) on the gauge field.

The integral over the field variables at all flow times $t>0$ 
can now be performed using the $\delta$-functions. Starting
with the fields at time $t=T$ and proceeding from there 
to the smaller times, the $\delta$-functions allow 
all fields to be eliminated recursively until the only
integration variables left are the fundamental fields $U$, 
$\psi$ and $\psibar$. In this process, the integrals over the 
$\delta$-functions for the gauge field variables give rise
to a product of Jacobians of the mapping $\Psu$, but 
these factors are canceled by the measure term in the action
(see appendix B).

The calculation thus shows that the correlation functions
in 4+1 dimensions of local fields that do not involve the 
Lagrange-multiplier fields
are exactly equal to the correlation functions
of the same fields
defined in 4 dimensions through the discrete flow 
equations and the QCD functional integral.

\subsection 5.4 Continuous time limit

At this point, the time-dependent fields in the correlation functions
are still the ones obtained by the Euler integration of the flow equations.
Since the associated
integration errors are of order $\eps$,
the continuous-time correlation functions are recovered 
when $\eps\to0$ (and, trivially, $T\to\infty$).
Moreover, the measure term
\equation{
  \Sms=\sum_{t=0}^{T-\eps}\sum_{x,\mu}
  \ln\det J\bigl(\rme^{-\eps g_0^2(\partial\Sw)(V,x,\mu)}\bigr)
  =\rmO(\eps)
  \enum
}
vanishes in this limit. 

In the following, the continuous-time formulation of the lattice
theory will normally be considered, but 
the functional integral in 4+1 dimensions is always assumed to be
defined through the limit $\eps\to0$ of
the discrete-time functional integral.

\section 6. On-shell O($\mib a$) improvement 

The lattice-spacing dependence 
of correlation functions involving the fields at non-zero flow time
is best discussed in the theory in 4+1 dimensions.
In this framework, the standard argumentation 
based on locality, power counting and symmetry
can be applied. 
The relevant Symanzik local effective theory
[\ref{SymanzikSeffI},\ref{SymanzikSeffII}] then
coincides with the continuum theory in 4+1 dimensions,
with O($a$) and higher-order terms in the lattice spacing $a$
added to the action and the local fields.
Moreover, on-shell improvement [\ref{OnShell}]
means that the cancellation of lattice-spacing effects
is limited to correlation functions 
at non-zero distances in 4+1 dimensions.

Although the theoretical discussion is more widely applicable,
the O($a$)-improved Wilson formulation of lattice QCD
[\ref{SW},\ref{SFimp}] is assumed from now on. 
Where possible, the same conventions
and notation as in ref.~[\ref{SFimp}] are used.

\subsection 6.1 Effective action

Since the flow equations in the lattice theory 
are classically O($a$) improved,
and since loop diagrams do not contribute to the correlation 
functions at asymptotically large flow times,
all O($a$) terms in the Symanzik effective action 
must be boundary terms, i.e.~local terms at flow time zero.
Furthermore, if the effective theory is to describe on-shell
quantities only,
as is the case here, many terms can be eliminated using the 
quark field equations,
the flow equations and the boundary conditions (2.1),(2.6). 
In the chiral limit, a possible choice 
of the O($a$) term in the effective action is then
\equation{
  a\int\rmd^4x\left\{c_1\psibar(x)\frac{i}{4}\sigma_{\mu\nu}F_{\mu\nu}(x)\psi(x)
  +c_2\lambdabar(0,x)\lambda(0,x)\right\},
  \enum
} 
where $F_{\mu\nu}(x)$ is the field tensor of the fundamental
gauge potential 
(here and below, $c_1,c_2,\ldots$ denote coefficients of the 
effective theory).

Quite many more terms are required to describe the 
O($a$) lattice-spacing effects at non-zero quark masses.
In the theory in 4 dimensions,
these were completely classified
by Bhattacharya et al.~[\ref{ImpNonD}].
The Symanzik effective action 
in 4+1 dimensions in addition includes a term
\equation{
  a\int\rmd^4x\left\{
  \lambdabar(0,x)(c_3M+c_4\kern0.5pt\tr M)\psi(x)+
  \psibar(x)(c_3M+c_4\kern0.5pt\tr M)\lambda(0,x)\right\}
  \enum
}
that contributes to correlation functions involving
the bulk fields, $M$ being the quark mass matrix in the effective theory. 
Its effect on the correlation functions is equivalent
to a renormalization
\equation{
  \chi(t,x)\to Z\chi(t,x),
  \quad
  \lambda(t,x)\to Z^{-1}\lambda(t,x),
  \enum
  \nexteq{2.5ex}
  Z=1+ac_3M+ac_4\kern0.5pt\tr M,
  \enum
}
of the bulk fermion (and anti-fermion) fields.

\subsection 6.2 Effective local fields

The form of the effective fields representing
the local lattice fields in the Symanzik theory
is constrained by the same general principles
that determine the form of the effective action.
In particular, the effective fields representing fields at positive
flow time have no O($a$) terms.

Such terms however occur in the case of the local
fields at flow time zero.
Considering again the theory with only massless quarks,
the effective fields representing the 
axial currents with flavour indices $r\neq s$, 
for example, are given by
\equation{
  (A_{\rm eff})_{\mu}^{rs}(x)
  =\psibar_r(x)\dirac{\mu}\dirac{5}\psi_s(x)+
   ac_5\partial_{\mu}\left\{\psibar_r(x)\dirac{5}\psi_s(x)\right\}
  \noenum
  \nexteq{2.0ex}
  {\phantom{(A_{\rm eff})_{\mu}^{rs}(x)=}}
  +ac_6\left\{\lambdabar_r(0,x)\dirac{\mu}\dirac{5}\psi_s(x)
  +\psibar_r(x)\dirac{\mu}\dirac{5}\lambda_s(0,x)
  \right\}+\ldots,
  \enum
}
where the ellipsis stands for terms of order $a^2$.
The associated effective pseudo-scalar densities,
\equation{
  (P_{\rm eff})^{rs}(x)=
  \psibar_r(x)\dirac{5}\psi_s(x)
  \noenum
  \noenum
  \nexteq{2.0ex}
  {\phantom{(P_{\rm eff})^{rs}(x)=}}
  +
  ac_7\left\{\lambdabar_r(0,x)\dirac{5}\psi_s(x)
  +\psibar_r(x)\dirac{5}\lambda_s(0,x)
  \right\}+\ldots,
  \enum
}
have a similar expansion.

The mass-dependent O($a$) corrections to these fields 
coincide with the ones in the theory in 4 dimensions
[\ref{ImpNonD}].
In general, the determination of the O($a$) 
terms however requires a careful consideration of a possible mixing
of fields and may lead to fairly complicated expressions.

\subsection 6.3\/ {\rm O($a$)\kern-0.5pt} improved lattice action

Following common practice, 
the mass-dependent O($a$) counterterms
are omitted in the improved action and are instead included
in the parameter and field renormalization factors [\ref{SFimp}].
The O($a$) counterterms
in 4+1 dimensions are then given by
\equation{
  \delta S_{\rm tot}=\sum_x\left\{\csw
  \psibar(x)\frac{i}{4}\sigma_{\mu\nu}\widehat{F}_{\mu\nu}(x)\psi(x)
  +\cfl\lambdabar(0,x)\lambda(0,x)\right\},
  \enum
}
where $\widehat{F}_{\mu\nu}(x)$ denotes the standard clover
expression for the gauge field tensor $F_{\mu\nu}(x)$ on the lattice.
While the first term in eq.~(6.7) (the Sheikholeslami--Wohlert term 
[\ref{SW},\ref{SFimp}]) is already required for the 
improvement of the theory in 4 dimensions, the one
proportional to the improvement
coefficient $\cfl(g_0)$
is needed to cancel the O($a$) lattice
effects in correlation functions involving the bulk fields.

Having specified the action,
the Wick contractions
of the basic fermion fields in the O($a$) improved theory
can be worked out straightforwardly 
(see appendix C).
In the continuous time limit, the contractions 
are practically the same
as in the continuum theory with the obvious modifications
(integrals over the space-time coordinates are replaced by
sums over lattice points, the quark propagator $S(x,y)$
by the inverse of the massive O($a$) improved lattice Dirac operator
and the kernel $K(t,x;s,y)$ by the fundamental solution of the 
quark flow equation on the lattice).
A special case is the contraction
\equation{
  \longwick{\chi(t,x)\chibar(s,y)}=\sum_{v,w}
  K(t,x;0,v)
  \left(S(v,w)-\cfl\delta_{vw}\right)
  K(s,y;0,w)^{\dagger}
  \enum 
}
of the time-dependent quark fields, which is the only one
depending on the improvement coefficient $\cfl$.
In particular, the contraction and thus the correlation functions
of the fields $\psi$ and $\psibar$ are independent of $\cfl$, as has to be
the case, since the theory in $4$ dimensions is already improved
through the inclusion of the Sheikholeslami--Wohlert term.

A perhaps puzzling aspect of eq.~(6.8) is the fact that the limit
\equation{
  \lim_{t\to0}\kern0.5pt\longwick{\chi(t,x)\chibar(s,y)}=
  \wick{\psi(x)\chibar(s,y)}-\cfl K(s,y;0,x)^{\dagger}
  \enum
}
differs from $\wick{\psi(x)\chibar(s,y)}$ by a 
term of order $a$, which is not just a contact term.
The subtractions required for
O($a$) improvement thus depend on whether the quark fields
in the correlation functions are at zero and 
non-zero flow time. This is actually not too surprising
since these fields also renormalize differently.
Both differences merely reflect the fact that 
the continuum limit of correlation
functions involving the time-dependent fields 
must be taken at fixed flow times given in physical units
and that the Symanzik effective theory in 4+1 dimensions
correctly describes the deviation of the lattice theory
from its continuum limit only when the latter is approached 
in this way.

\subsection 6.4 Renormalized improved fields

As usual, the additively renormalized bare quark masses 
$\mq{r}$ are defined by
\equation{
  \mq{r}=m_{0,r}-m_{\rm c},
  \enum
}
where $m_{\rm c}$ denotes the critical bare mass in the theory 
with mass-degenerate quarks.
It is also helpful to introduce
the associated subtracted bare mass matrix $\Mq$ and the combinations
\equation{
   \mq{rs}=\frac{1}{2}\left(\mq{r}+\mq{s}\right)
   \enum
}
of quark masses.

The lattice fermion fields at positive flow time require multiplicative 
renormalization and O($a$) improvement by a mass-dependent factor
(cf.~subsect.~6.1).
Quark and anti\-quark fields 
are scaled with the same factor,
\equation{
  \rens{\chi}{r}=\{Z_{\chi}(1+b_{\chi}\mq{r}+
  \bar{b}_{\chi}\tr\Mq)\}^{1/2}\chi_r,
  \enum
}
while the Lagrange-multiplier fields $\lambda,\lambdabar$ 
are renormalized with the inverse of the same factor%
\kern1pt\footnote{$\dagger$}{\footnotefont%
Following refs.~[\ref{SFimp},\ref{ImpNonD}],
$b_X$ and $\bar{b}_X$ generically denote improvement
coefficients multiplying mass-dependent O($a$) 
counterterms.}.
Time-dependent composite fields like the 
pseudo-scalar density 
\equation{
  \rens{P}{t}^{rs}=Z_{\chi}(
  1+b_{\chi}\mq{rs}+\bar{b}_{\chi}\tr\Mq)
  P_t^{rs}
  \enum
}
then renormalize according to their quark content.

Starting from the bare fields
\equation{
  A_{\mu}^{rs}(x)=\psibar_r(x)\dirac{\mu}\dirac{5}\psi_s(x),
  \enum
  \nexteq{2.5ex}
  \tilde{A}^{rs}_{\mu}(x)=
  \lambdabar_r(0,x)\dirac{\mu}\dirac{5}\psi_s(x)
  +\psibar_r(x)\dirac{\mu}\dirac{5}\lambda_s(0,x),
  \enum
  \nexteq{2.5ex}
  P^{rs}(x)=\psibar_r(x)\dirac{5}\psi_s(x),
  \enum
  \nexteq{2.5ex}
  \tilde{P}^{rs}(x)=
  \lambdabar_r(0,x)\dirac{5}\psi_s(x)
  +\psibar_r(x)\dirac{5}\lambda_s(0,x),
  \enum
}
the improved flavour non-singlet axial current and density are
given by
\equation{
  \imps{A}{\mu}^{rs}=A^{rs}_{\mu}
  +\ca\ring{\partial}_{\mu}P^{rs}
  +\cta\tilde{A}^{rs}_{\mu},
  \enum
  \nexteq{2.5ex}
  \imp{P}^{rs}=P^{rs}+\ctp\tilde{P}^{rs},
  \enum
}
and these are renormalized according to [\ref{SFimp},\ref{ImpNonD}]
\equation{
  \rens{A}{\mu}^{rs}=\ZA(1+b_A\mq{rs}+\bar{b}_A\tr\Mq)
  \imps{A}{\mu}^{rs},
  \enum
  \nexteq{2.5ex}
  \ren{P}^{rs}=\ZP(1+b_P\mq{rs}+\bar{b}_P\tr\Mq)
  \imp{P}^{rs}.
  \enum
}
The new improvement coefficients, $\cta$ and $\ctp$, are required for
the O($a$) improvement of correlation functions involving the fermion
fields at non-zero flow time, but all other coefficients already
occur in the theory in four dimensions [\ref{SW}--\ref{ImpNonD}].
At tree-level of perturbation theory,
\equation{
  \cfl=\frac{1}{2},
  \qquad
  \cta=\ctp=-\frac{1}{2},
  \qquad
  b_{\chi}=1.
  \enum
}
All coefficients $\bar{b}_X$ are, incidentally, of order $g_0^4$
[\ref{ImpNonD}].

\subsection 6.5 Improvement and renormalization of the field $\Ptax^{rs}$

In the chiral symmetry relations derived in 
sect.~4, the field $\Ptax^{rs}$ plays a prominent r\^ole.
The improvement and renormalization is a bit complicated 
in this case and is therefore discussed separately from the
other fields.

As already noted in subsect.~4.2, the field
can only mix with 
local composite fields that include at least one Lagrange-multiplier 
field. Charge conjugation, the lattice symmetries and the flavour symmetry
then imply that the field is multiplicatively renormalizable. 
There are, however, several fields that can mix 
with the density at order $a$,
among them two fields
\equation{
  \hat{P}^{rs}(x)=\lambdabar_r(0,x)\dirac{5}\lambda_s(0,x),
  \enum
  \nexteq{2.5ex}
  Q^{rs}(x)=\lambdabar_r(0,x)\dirac{5}\psi_s(x)-
  \psibar_r(x)\dirac{5}\lambda_s(0,x),
  \enum
}
that have not appeared before.
Inspection then shows that 
a possible choice of the improved and renormalized densities is
\equation{
  \imp{\Ptax}^{rs}=
  \Ptax^{rs}+\hat{c}_A
  \ring{\partial}_{\mu}
  \tilde{A}_{\mu}^{rs}+\hat{c}_P\hat{P}^{rs},
  \enum
  \nexteq{2.5ex}
  \ren{\Ptax}^{rs}=
  \tilde{Z}_P\bigl\{
  (1+b_{\Ptax}\mq{rs}+\bar{b}_{\Ptax}\tr\Mq)
  \imp{\Ptax}^{rs}
  +\hat{b}_{\Ptax}(\mq{r}-\mq{s})Q^{rs}\bigr\},
  \enum
}
where, as usual, the field equations were used to reduce the
number of terms.
The fields on the right of these equations 
are distinguished by their symmetry properties or content in
Lagrange-multiplier fields.
They are all multiplicatively renormalizable and the contributions
of the counterterms proportional to $\tilde{A}_{\mu}^{rs}$, $\hat{P}^{rs}$ and
$Q^{rs}$ in on-shell correlation functions are therefore of order $a$.

The expression (6.25),(6.26) for the renormalized improved density
can be simplified by noting that the relations
\equation{
  \sum_x\langle\Ptax^{rs}(x)P^{sr}_t(y)\rangle
  =\langle S_t^{rr}(y)\rangle+\langle S_t^{ss}(y)\rangle
  +2\cfl\sum_x\langle\hat{P}^{rs}(x)P^{sr}_t(y)\rangle,
  \enum
  \nexteq{2.5ex}
  \sum_x\langle Q^{rs}(x)P^{sr}_t(y)\rangle
  =\langle S_t^{ss}(y)\rangle-\langle S_t^{rr}(y)\rangle,
  \enum
}
hold exactly, for any $t>0$, as a consequence of
the form of the Wick contractions of the fermion fields. 
On the other hand, as discussed
in subsect.~4.2, the renormalization
constants can be (and are to be) chosen so that the normalization 
condition (4.15) is satisfied in the continuum limit.
Since the correlation functions in eq.~(4.15) converge to the
continuum limit with a rate proportional to $a^2$,
the comparison with the unrenormalized identities, eqs.~(6.27),(6.28),
then shows that
\equation{
  \tilde{Z}_P=1,
  \enum
  \nexteq{2.5ex}
  \hat{c}_P=-2\cfl,
  \quad
  b_{\Ptax}=\bar{b}_{\Ptax}=0,
  \quad
  \hat{b}_{\Ptax}=-\frac{1}{2}b_{\chi}.
  \enum
}
The renormalized density 
\equation{
  \ren{\Ptax}^{rs}=
  \Ptax^{rs}+
  \hat{c}_A
  \ring{\partial}_{\mu}
  \tilde{A}_{\mu}^{rs}-2\cfl\hat{P}^{rs}
  -b_{\chi}\frac{1}{2}(\mq{r}-\mq{s})Q^{rs}
  \enum
}
thus assumes a fairly simple form,
in which $\hat{c}_A$ is the only new improvement coefficient.

\subsection 6.6 PCAC relation on the lattice

In the continuum limit, the renormalized PCAC relation (4.14) holds 
provided the renormalization constants $Z_A,Z_P$ and the renormalized
quark masses $\mR{r}$ are chosen appropriately.
If also all improvement coefficients are properly tuned, this
implies
\equation{
  \langle\partial\ren{A}^{rs}(x)\rens{P}{t}^{sr}(y)\rangle
  \noenum
  \nexteq{2.0ex}
  \qquad
  =(\mR{r}+\mR{s})\langle\ren{P}^{rs}(x)\rens{P}{t}^{sr}(y)\rangle
  -\langle\ren{\Ptax}^{rs}(x)\rens{P}{t}^{sr}(y)\rangle
  +\rmO(a^2),
  \enum
}
where, following the tradition, the divergence of the 
improved axial current is taken to be
\equation{
  \partial\ren{A}^{rs}=\ZA\left(1+b_A\mq{rs}+\bar{b}_A\tr\Mq\right)
  \partial\imp{A}^{rs},
  \enum
  \nexteq{2.5ex}
  \partial\imp{A}^{rs}=
  \ring{\partial}_{\mu}\{A^{rs}_{\mu}+
  \cta\tilde{A}^{rs}_{\mu}\}
  +\ca\drvstar{\mu}\drv{\mu}P^{rs}.
  \enum
}
A technical detail worth emphasizing here again is the fact that
the PCAC relation (6.32) holds at all $x$, including $x=y$, as long as 
the flow time $t$ is set to a positive value given in physical units.

\section 7. Calculation of correlation functions

In numerical lattice QCD,
correlation functions involving fields 
at non-zero flow time can be computed
following the steps usually taken in the 
case of ordinary hadronic correlation functions.
The fact that the flow equations must be solved at 
some point nevertheless represents a complication
that needs to be carefully considered.
For illustration, the details are worked out in this section
for two representative cases.

\subsection 7.1 Pseudo-scalar two-point function

In practice one is interested in the
correlation function
\equation{
  \langle \Pax^{rs}(x)\Pax_t^{sr}(y)\rangle=
  -\sum_{v,w}\langle\tr\{
  [K(t,y;0,v)S(v,x)_{ss}]^{\dagger}K(t,y;0,w)S(w,x)_{rr}\}\rangle
  \enum
}
at vanishing spatial momentum and computes
its average over a lattice $\Gamma$ of source points,
using random source fields [\ref{MichaelPeisa}],
in order to reduce the statistical errors.

If $x$ is taken to be
the source point, the averaging amounts to the substitution
\equation{
  P^{rs}(x)\to {1\over N_{\Gamma}N_s}
  \sum_{k=1}^{N_s}
  (\psibar_r,\eta_k)(\eta_k,\dirac{5}\psi_s),
  \enum
}
where $N_{\Gamma}$ is the number of points in $\Gamma$ and
\equation{
  \eta_k(x),\quad k=1,\ldots,N_s,
  \enum
}
are randomly chosen complex spinor fields on $\Gamma$ with mean zero
and variance
\equation{
  \langle\eta_k(v)\eta_l(w)^{\dagger}\rangle=
  \cases{\delta_{kl}\delta_{vw} & if $v,w\in\Gamma$,\cr
  \noalign{\vskip1.0ex}
  0 & otherwise.\cr}
  \enum
}
In a QCD simulation,
the computation of the 
averaged correlation function at zero spatial momentum,
\equation{
  {1\over N_{\Gamma}}\sum_{x\in\Gamma}\sum_{\vec{y}}
  \langle \Pax^{rs}(x)\Pax_t^{sr}(y)\rangle=
  -{1\over N_{\Gamma}N_s}\sum_{k=1}^{N_s}
  \sum_{\vec{y}}\langle\phi_{k,s}(t,y)^{\dagger}\phi_{k,r}(t,y)\rangle,
  \enum
}
then amounts to calculating the functions
\equation{
  \phi_{k,h}(t,y)=\sum_{w,x}
  K(t,y;0,w)S(w,x)_{hh}\eta_k(x)
  \enum
}
for a representative ensemble of gauge fields,
all $k=1,\ldots,N_s$ and $h=r,s$.

For a given gauge field,
the computation of $\phi_{k,h}(t,y)$ proceeds in two steps,
where one first calculates $\phi_{k,h}(0,w)$ by solving
the lattice Dirac equation with mass $m_{0,h}$ and
source field $\eta_k(x)$. The calculated field must then 
be evolved in flow time from time $0$ to time $t$.
In the direction of increasing flow time,
the flow equations
are numerically stable and the solution can easily 
be obtained, with negligible
integration errors, using a Runge--Kutta integrator
(appendix D).

\subsection 7.2 Chiral condensate

In the case of the time-dependent quark condensate,
\equation{
  \langle S_t^{rr}(x)\rangle=
  -\sum_{v,w}\langle
  \tr\{K(t,x;0,v)\left[S(v,w)_{rr}-\cfl\delta_{vw}\right]
  K(t,x;0,w)^{\dagger}\}\rangle,
  \enum
}
an averaging over the position $x$ is again desirable. 
Introducing random source fields as above, one is then
led to the representation
\equation{
  {1\over N_{\Gamma}}\sum_{x\in\Gamma}
  \langle S_t^{rr}(x)\rangle=
  -{1\over N_{\Gamma}}
  \sum_{v,w}\langle
  \xi_k(t;0,v)^{\dagger}\left[S(v,w)_{rr}-\cfl\delta_{vw}\right]
  \xi_k(t;0,w)\rangle
  \enum
}
in terms of the fields
\equation{
  \xi_k(t;s,w)=\sum_x K(t;x;s,w)^{\dagger}\eta_k(x)
  \enum
}
at flow time $s=0$. 

The computation of these fields requires 
the {\it adjoint flow equation}
\equation{
  (\partial_s+\Delta)\xi_k(t;s,w)=0
  \enum
}
to be solved from time $s=t$ (where $\xi_k$ coincides
with $\eta_k$) to $s=0$. A straightforward Runge--Kutta
integration should not be used here, because
the Laplacian $\Delta$ is the one in presence
of the gauge field determined by the gradient flow at time $s$,
which would therefore have to be evolved backward in time,
i.e.~in the unstable direction.
As explained in appendix E, this difficulty can be overcome
through a hierarchical scheme that avoids the backward integration
of the gauge field.

\section 8. Strategy for the computation of $\mib\ZA$ and ${\mib c}_{\bf fl}$

The PCAC relation (6.32) holds provided 
the relevant renormalization constants and improvement coefficients 
are chosen appropriately. 
As explained below,
the constant $\ZA$ and the 
coefficient $\cfl$ can conversely be determined,
up to terms of order $a^2$ and $a$, respectively,
by requiring the relation to be satisfied exactly at
two points in the space of kinematical parameters.

The computation through numerical simulation 
of $\ZA$ and $\cfl$ along these lines
assumes that the improvement coefficients 
$\csw$ and $\ca$ have already been calculated.
Periodic boundary conditions should be
imposed in the space directions, but apart from this no
further assumptions are made. In particular,
the boundary conditions in time do not need to be
specified, since the
computed values of $\ZA$ and $\cfl$ 
depend on them only
at the level of the lattice-spacing effects.

\subsection 8.1 Integrated form of the PCAC relation

The O($a$) improvement unfortunately leads to an inflation
of terms when the renormalized PCAC relation is expanded
in the correlation functions of the unimproved bare fields.
A slight simplification can however be achieved by summing
the equation over an interval of time $x_0$ 
around $y_0$ and over
all points $\vec{x}$ on the spatial lattice. 
One of the bare correlation functions entering the 
relation is then
\equation{
  \CPP{rs}(t,d)=\sum_{x_0=y_0-d}^{y_0+d}\sum_{\vec{x}}
  \langle\Pax^{rs}(x)\Pax_t^{sr}(y)\rangle,
  \enum
}
and there are $6$ further such correlation functions in which
$P^{rs}$ is replaced by
\equation{
  \tilde{P}^{rs},\hat{P}^{rs},Q^{rs},\ring{\partial}_{\mu}A_{\mu}^{rs},
  \ring{\partial}_{\mu}\tilde{A}_{\mu}^{rs}
  \kern0.5em\hbox{or}\kern0.5em
  \drvstar{\mu}\drv{\mu}P^{rs}.
  \enum
}
In the case of the last three fields in this list, 
the sum over $x_0$ reduces to 
a sum of terms at the boundary of the summation range as in 
\equation{
  \CAP{rs}(t,d)=
  \frac{1}{2}\sum_{\vec{x}}\bigl\{
  \langle(\left.A^{rs}_0(x)\right|_{x_0=y_0+d+1}+
          \left.A^{rs}_0(x)\right|_{x_0=y_0+d})
  P^{sr}_t(y)\rangle
  \noenum
  \nexteq{1.0ex}
  {\phantom{\CAP{rs}(t,d)=
  \frac{1}{2}\sum_{\vec{x}}}}\kern-0.65em
  -\langle(\left.A^{rs}_0(x)\right|_{x_0=y_0-d-1}+
           \left.A^{rs}_0(x)\right|_{x_0=y_0-d})
  P^{sr}_t(y)\rangle
  \bigr\},
  \enum
}
for example.

For notational convenience, it is now helpful 
to introduce the abbreviations
\equation{
  \ZAh{rs}=\ZA\left(1+b_A\mq{rs}+\bar{b}_A\tr\Mq\right),
  \enum
  \nexteq{2.5ex}
  \ZPh{rs}=\ZP\left(1+b_P\mq{rs}+\bar{b}_P\tr\Mq\right).
  \enum
}
Defining the bare current-quark mass sums through
\equation{
  m_{rs}={\ZPh{rs}\over\ZAh{rs}}(\mR{r}+\mR{s}),
  \enum
}
the integrated PCAC relation then becomes
\equation{
  \ZAh{rs}\bigl\{\CAP{rs}+
  \tilde{c}_A\CAtP{rs}+c_A\CdPP{rs}
  -m_{rs}\bigl(\CPP{rs}+\tilde{c}_P\CPtP{rs}\bigr)\bigr\}
  \noenum
  \nexteq{2.0ex}
  \qquad+\CPtP{rs}+\hat{c}_A\CAtP{rs}-2\cfl\CPhP{rs}
  -b_{\chi}\frac{1}{2}(\mq{r}-\mq{s})\CQP{rs}=\rmO(a^2).
  \enum
}
This equation still looks fairly complicated, but one should keep in mind
that it is a linear relation among computable correlation functions,
which must hold for all flow times $t>0$, time ranges
$d\geq0$ and quark flavours
$r\neq s$. The unknown coefficients are thus strongly
constrained by the identity.

At values of $d$ larger than the smoothing range 
$\sqrt{8t}$ of the gradient flow, $\CAtP{rs}$ falls off like a Gaussian
and rapidly becomes negligible with respect to the other terms in
eq.~(8.7). Moreover, the sum of the terms in the curly bracket 
as well as all other terms in the equation 
are then practically independent of $d$.
Without further notice, only this 
range of $d$ is considered in the following and the
two terms proportional to $\CAtP{rs}$ are dropped.

\subsection 8.2 Up-down flavour channel

In QCD with mass-degenerate up and down quarks,
the PCAC relation in the up-down channel assumes the form
\equation{
  \ZAh{ud}\bigl\{\CAP{ud}+c_A\CdPP{ud}-m_{ud}\CPP{ud}\bigr\}
  -2\cfl\CPhP{ud}
  =-\bigl(1-\ZAh{ud}\tilde{c}_Pm_{ud}\bigr)\CPtP{ud}
  \enum
}
(for simplicity, the O($a^2$) error term is dropped from now on).
The bare mass combination $m_{ud}$ can be determined as usual
through the PCAC relation at vanishing flow time
and may be assumed to be known at this point.
By computing the correlation functions in eq.~(8.8) at 
two values of the flow time $t$, one then obtains
two linear equations for the ratios
\equation{
  {\ZAh{ud}\over1-\ZAh{ud}\tilde{c}_Pm_{ud}}
  \quad\hbox{and}\quad
  {\cfl\over1-\ZAh{ud}\tilde{c}_Pm_{ud}}.
  \enum
}
Clearly, as long as the chosen flow times are
in the scaling range and are held fixed in physical units,
the calculated ratios do not depend on them,
up to terms of order $a^2$
and $a$, respectively. 

At quark masses close to their physical values, and lattice spacings
$a\leq0.1$ fm,
the factor $1-\ZAh{ud}\tilde{c}_Pm_{ud}$ differs
from unity by a fraction of a percent.
An estimate of the size of this correction
term can be obtained by inserting the known value of $m_{ud}$ and 
the tree-level value (6.22) of $\tilde{c}_{P}$.
$\ZAh{ud}$ and $\cfl$ can thus be computed up to this
small correction factor. Note that the bare pion decay constant
determined through the vacuum-to-pion matrix element of the improved
axial current is to be renormalized with the
mass-dependent factor $\ZAh{ud}$ rather than $\ZA$. The difference
is however small and can be estimated using the 
tree-level value $b_A=1$.

\subsection 8.3 Up-strange flavour channel

The additional term that appears in the PCAC relation
\equation{
  \ZAh{us}\bigl\{\CAP{us}+c_A\CdPP{us}-m_{us}\CPP{us}\bigr\}
  -2\cfl\CPhP{us}-b_{\chi}\frac{1}{2}(\mq{u}-\mq{s})\CQP{us}
  \noenum
  \nexteq{2.0ex}
  \qquad
  =-\bigl(1-\ZAh{us}\tilde{c}_Pm_{us}\bigr)\CPtP{us}
  \enum
}
in this channel is proportional to the square of the mass 
difference $\mq{u}-\mq{s}$ and is therefore 
likely to be negligible in practice.
To be able to determine the ratios (8.9) (with $ud$ replaced by $us$),
eq.~(8.10) must otherwise be evaluated at
three values of the flow time $t$.

For $a\leq0.1$ fm and physical quark masses,
the correction factor $1-\ZAh{us}\tilde{c}_Pm_{us}$ 
may deviate from unity by up to one percent or so.
The comparison with the results obtained in the $ud$ channel,
\equation{
  {\ZAh{us}(1-\ZAh{ud}\tilde{c}_Pm_{ud})
  \over\ZAh{ud}(1-\ZAh{us}\tilde{c}_Pm_{us})}
  =1+b_A\frac{1}{2}(\mq{s}-\mq{d})+\ZA\tilde{c}_P(m_{us}-m_{ud}),
  \enum
}
unfortunately only allows a certain linear combination of the coefficients 
$b_A$ and $\tilde{c}_P$ to be estimated.

\section 9. Sample calculation in 2+1 flavour QCD

The aim in the following paragraphs is to demonstrate
the feasibility of the proposed strategies for the 
computation of the renormalization constant $\ZA$, 
the improvement coefficient $\cfl$ and the low-energy constants
$\Sigma$ and $F$. A single simulation run suffices for this
test and no attempt is made to actually compute the 
low-energy constants,
as this would require the data to be extrapolated 
to the infinite-volume, chiral and continuum limit.

\subsection 9.1 Simulation details

The calculations reported below are based on a
recent simulation of QCD with 2+1 flavours of quarks [\ref{TMopenQCD}].
In this run (labeled $I_1$ in ref.~[\ref{TMopenQCD}]), 
a $64\times32^3$ lattice with open boundary conditions
in time [\ref{openQCD}]
was simulated at a point in parameter space, where
the lattice spacing is estimated to be $0.090$ fm
[\ref{AokiEtAlI},\ref{AokiEtAlII}] and where 
the pion and kaon masses are about 203 and 520 MeV,
respectively [\ref{TMopenQCD}].
The physical size of the lattice is thus 
such that the finite-volume effects on the calculated 
meson masses and matrix elements are expected to be small.

For the Iwasaki gauge action [\ref{Iwasaki}]
and the standard O($a$)-improved
Wilson--Dirac operator [\ref{SW},\ref{SFimp}] used
in the simulation, the coefficients 
$\csw$ and $\ca$ are known
non-perturbatively [\ref{AokiEtAlIII},\ref{KanekoEtAl}].
A representative ensemble of 150 gauge-field configurations,
separated by $9.6$ units of molecular-dynamics time,
was produced in the run.
In order to suppress
any residual autocorrelation effects,
the statistical errors quoted in this section
were estimated by 
dividing the measurement data series for the primary observables
into bins of 5 consecutive measurements
and by applying the jackknife method to the binned data.

\topinsert
\vbox{
\vskip0.0cm
\centerline{\epsfxsize=10.5 true cm\epsfbox{plots/wi.eps}}
\vskip0.3cm
\figurecaption{%
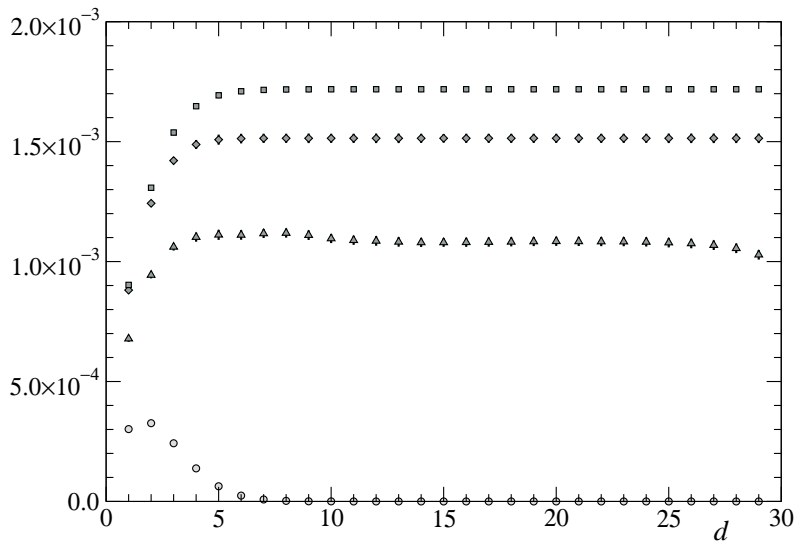
Combinations of correlation functions appearing
in the PCAC relation (8.8),
plotted as a function of the summation range $d$.
The linear combinations of correlation functions shown
are the ones multiplying the renormalization constant $\ZAh{ud}$ (triangles),
the improvement coefficient $\cfl$ (diamonds)
and the factor $1-\ZAh{ud}\ctp m_{ud}$ (squares)
at flow time $t=3.86$.
For $d\geq9$, the function 
$\CAtP{ud}$ (circles) is many orders of magnitude smaller than the other 
functions and can be safely neglected (cf.~subsect.~8.1).
All quantities are given in lattice units.
}
}
\endinsert

\subsection 9.2 Computation of $\ZAh{rs}$ and $\cfl$

The values $2.47,\,3.86,\,5.56$ and $7.56$
of the flow time $t$ in lattice units, where
the correlation functions entering the PCAC relation (8.7)
were computed, correspond to smoothing ranges of the gradient
flow equal to $0.4,\,0.5,\,0.6$ and $0.7$ fm, respectively.
In order to keep away from the boundaries of the lattice
in the time direction, the
time-dependent field $P_t^{rs}(y)$ was inserted
in the middle of the lattice. The 
correlation functions were then averaged over space translations
using 12 random source fields at time $y_0$.

All correlation functions in the PCAC relation
can easily be obtained with small statistical errors over
the whole range of the summation window $d$.
As expected, the sum of correlation functions multiplying
the axial-current renormalization constant and the other
correlation functions reach a plateau
at values of $d$ larger than 2 or 3 times the smoothing range 
of the gradient flow (see fig.~1 for a representative case;
the slight bending down of the 
function multiplying $\ZAh{ud}$ close to the 
boundary of the lattice is a lattice-spacing effect).

Following the strategy outlined in sect.~8, the ratios (8.9)
are now determined by requiring the PCAC relation (8.8) to hold exactly 
at two values $t_1,t_2$ of $t$.
The choice of the summation window $d$ has very little influence 
on the calculated ratios as long as one stays within
the plateau range of the correlation functions.
Setting $d=18$ then leads to the results
quoted in the second and third column of table~1,
where the factor $1-\ZAh{ud}\ctp m_{ud}$ was estimated
using the tree-level value (6.22) of the coefficient $\ctp$.
As anticipated, this correction is very small (about $0.2\%$).

In the up-strange flavour channel, there are two sources of 
O($a$) mass corrections. Proceeding as above, the correction
coming from the factor $1-\ZAh{us}\ctp m_{us}$
is estimated to be approximately $1.3\%$, 
while the one deriving from the 
term in eq.~(8.10) proportional to $b_{\chi}$
changes the results by about $0.2\%$.
As can be seen from table~1,
there is a remarkable consistency among the
values of $\cfl$ obtained in the two flavour channels.
There is also little dependence on the flow
times and the difference of the renormalization constants
$\ZAh{ud}$ and $\ZAh{us}$ might very well be explained by
the mass-dependent factor in eq.~(8.4). The O($a$) improvement
thus works out very well, with residual 
lattice-spacing effects at the level of a fraction of a percent 
in the case of the renormalization constant.

\topinsert
\newdimen\digitwidth
\setbox0=\hbox{\rm 0}
\digitwidth=\wd0
\catcode`@=\active
\def@{\kern\digitwidth}
\tablecaption{Simulation results for $\ZAh{rs}$ and $\cfl$} 
\vskip-3.0ex

$$\vbox{\settabs\+&%
                  xxxxxxxxxxx&xxx&
                  xxxxxxxxxxxx&xx&
                  xxxxxxxxxxxx&xx&
                  xxxxxxxxxxxx&xx&
                  xxxxxxxxxxxx&\cr
\thicktablerule
\vskip1.2ex
                \+& \hfill $ t_1,t_2$\hfill
                 && \hfill $\ZAh{ud}$\hfill
                 && \hfill $\cfl$\hfill
                 && \hfill $\ZAh{us}$\hfill
                 && \hfill $\cfl$\hfill
                 &\cr
\vskip1.0ex
\thintablerule
\vskip1.2ex
  \+& \hfill $2.47,3.86$\hfill
  &&  \hfill $0.7929(87)$\hfill
  &&  \hfill $0.5703(11)$\hfill
  &&  \hfill $0.8096(41)$\hfill
  &&  \hfill $0.56867(60)$\hfill
  &\cr
\vskip0.3ex
  \+& \hfill $2.47,5.56$\hfill
  &&  \hfill $0.7904(85)$\hfill
  &&  \hfill $0.5712(14)$\hfill
  &&  \hfill $0.8073(40)$\hfill
  &&  \hfill $0.56990(72)$\hfill
  &\cr
\vskip0.3ex
  \+& \hfill $2.47,7.56$\hfill
  &&  \hfill $0.7895(85)$\hfill
  &&  \hfill $0.5716(16)$\hfill
  &&  \hfill $0.8067(39)$\hfill
  &&  \hfill $0.57025(83)$\hfill
  &\cr
\vskip0.3ex
  \+& \hfill $3.86,5.56$\hfill
  &&  \hfill $0.7885(85)$\hfill
  &&  \hfill $0.5734(22)$\hfill
  &&  \hfill $0.8054(40)$\hfill
  &&  \hfill $0.5726(10)@$\hfill
  &\cr
\vskip0.3ex
  \+& \hfill $3.86,7.56$\hfill
  &&  \hfill $0.7883(85)$\hfill
  &&  \hfill $0.5735(26)$\hfill
  &&  \hfill $0.8055(39)$\hfill
  &&  \hfill $0.5724(12)@$\hfill
  &\cr
\vskip0.3ex
  \+& \hfill $5.56,7.56$\hfill
  &&  \hfill $0.7882(86)$\hfill
  &&  \hfill $0.5738(36)$\hfill
  &&  \hfill $0.8056(38)$\hfill
  &&  \hfill $0.5723(16)@$\hfill
  &\cr
\vskip1.2ex
\thicktablerule
\vskip1.0ex
\+{\footnotefont The values of $\cfl$ in the 3rd and 5th column
were obtained in the $ud$ and $us$ channel, respectively}\cr
}
$$
\vskip-2.0ex
\endinsert

The results for the renormalization constant 
$\ZAh{ud}$ quoted in table~1 are in agreement
with the value $\ZA=0.781(20)$ previously
obtained in ref.~[\ref{AokiEtAlIV}]
using a method based on the Schr\"odinger functional
[\ref{ZASFI},\ref{ZASFII}]. Note that the improvement coefficient 
$\cfl$ turns out to be close to its tree-level value (6.22),
although there is no very good reason for this to be so
at the gauge coupling considered.

\subsection 9.3 Computation of the chiral condensate

Having determined the improvement coefficient $\cfl$, the unrenormalized
time-de\-pen\-dent 
condensate $\Sigma^{uu}_t$ can be computed straightforwardly
as described in subsect.~7.2.
The source time $x_0$ is taken to 
be in the middle of the lattice in order to minimize 
the excited-states contributions to the expectation value of
the scalar density. 
Using 12 random source fields, and evaluating
$\cfl$ at $t_1,t_2=3.86,5.56$, the calculation yields the results
quoted in the second column of table~2. 
In the range of flow time considered,
the time-dependent 
condensate is thus seen to decrease, roughly like $1/t$, 
when $t$ increases. The statistical error follows this
evolution and is, in any case, encouragingly small.

\topinsert
\newdimen\digitwidth
\setbox0=\hbox{\rm 0}
\digitwidth=\wd0
\catcode`@=\active
\def@{\kern\digitwidth}
\tablecaption{Simulation results for 
$\Sigma^{uu}_t$ and $G^{ud}_t$ in lattice units} 
\vskip-3.0ex
$$\vbox{\settabs\+&%
                  xxxxxxx&xx&
                  xxxxxxxxxxxxx&xx&
                  xxxxxxxxxxxxx&xx&
                  xxxxxxxxxxxxx&xx&
                  xxxxxxxxxxxxx&\cr
\thicktablerule
\vskip1.2ex
                \+& \hfill $t$\hfill
                 && \hfill $\Sigma^{uu}_t$\hfill
                 && \hfill $G^{ud}_t$\hfill
                 && \hfill $\Sigma^{uu}_tG^{ud}/G^{ud}_t$\hfill
                 && \hfill $\Sigma^{uu}_t/G^{ud}_t$\hfill
                 &\cr
\vskip1.0ex
\thintablerule
\vskip1.2ex
  \+& \hfill $2.47$\hfill
  &&  \hfill $0.0006277(95)$\hfill
  &&  \hfill $0.01484(23)@$\hfill
  &&  \hfill $0.003962(61)$\hfill
  &&  \hfill $0.04230(85)$\hfill
  &\cr
\vskip0.3ex
  \+& \hfill $3.86$\hfill
  &&  \hfill $0.0004251(58)$\hfill
  &&  \hfill $0.01028(16)@$\hfill
  &&  \hfill $0.003872(55)$\hfill
  &&  \hfill $0.04134(81)$\hfill
  &\cr
\vskip0.3ex
  \+& \hfill $5.56$\hfill
  &&  \hfill $0.0002911(36)$\hfill
  &&  \hfill $0.00720(11)@$\hfill
  &&  \hfill $0.003785(51)$\hfill
  &&  \hfill $0.04040(78)$\hfill
  &\cr
\vskip0.3ex
  \+& \hfill $7.56$\hfill
  &&  \hfill $0.0002017(23)$\hfill
  &&  \hfill $0.005092(80)$\hfill
  &&  \hfill $0.003711(48)$\hfill
  &&  \hfill $0.03962(76)$\hfill
  &\cr
\vskip1.2ex
\thicktablerule
}
$$
\vskip-2.0ex
\endinsert

To be able to relate the time-dependent condensate to the 
quark condensate $\Sigma$ in the chiral limit,
the vacuum-to-pion matrix elements 
\equation{
  \Gpi=\ZP(1+b_P\mq{ud}+\bar{b}_P\tr\Mq)\kern1pt G^{ud},
  \enum
  \nexteq{3.0ex}
  \Gpit=Z_{\chi}(1+b_{\chi}\mq{ud}+\bar{b}_{\chi}\tr\Mq)\kern1pt G^{ud}_t,
  \enum
}
need to be computed as well. Actually, since only the ratio
\equation{
 {\rens{\Sigma}{t}\over\Gpit}=
 {\Sigma^{uu}_t\over G^{ud}_t}
 \enum
}
appears in eqs.~(4.24),(4.25), it suffices to calculate $\Gpi$ 
and the unrenormalized matrix element $G^{ud}_t$. The 
latter can be determined from
the pseudo-scalar correlation function (7.1) at 
large time separations $|x_0-y_0|$ in the same way as
the bare matrix element $G^{ud}$ at vanishing flow time 
is usually extracted from the ordinary pseudo-scalar two-point function.
Setting $x_0$ to a value next to the boundaries of the lattice, 
there is in all cases a wide range in time $y_0$, 
where the one-pion intermediate states dominate
and the desired matrix elements can be easily determined.

Similarly to the time-dependent condensate, the matrix element
$G^{ud}_t$ is a monotonically decreasing function of $t$.
The ratios listed in the last two columns of table~2 however
vary much less with $t$, consistently with
the fact that the flow-time dependence of the ratios must disappear
in the chiral limit.

\subsection 9.4 Remarks

{\it Conversion to physical units.}
The spacing 
of the simulated lattice,
$a=0.08995(40)$ fm, was determined by PACS-CS 
through a computation of the mass of the $\Omega$ baryon
[\ref{AokiEtAlII}]. Using step scaling and the Schr\"odinger functional,
PACS-CS also calculated the renormalization constant
$\ZP=0.580(21)$ [\ref{AokiEtAlIV}] 
that relates the bare matrix element $G^{ud}$ to
$\Gpi$ in the $\MSbar$ 
scheme at 2 GeV\kern1pt\footnote{$\dagger$}{\footnotefont%
A combination of table entries quoted in ref.~[\ref{AokiEtAlIV}],
with errors added in quadrature,
was required to obtain the value of $\ZP$ given here.}. 
At $t=3.86$, for example, the results
\equation{
  {\rens{\Sigma}{t}\Gpi\over\Gpit}=[287(4)\,\MeV]^3,
  \enum
  \nexteq{2.5ex}
  {\rens{\Sigma}{t}\over\Gpit}=91(2)\,\MeV,
  \enum
}
are then obtained, where the O($a$) mass correction in eq.~(9.1) was
neglected (the error in eq.~(9.4) is anyway dominated by 
the error of $\ZP$). Clearly, while these ratios may be close
to the condensate $\Sigma$ and decay constant $F$ in the chiral limit,
simulations of a range of lattices will be required
to be able to determine $\Sigma$ and $F$ with controlled 
systematic errors.

\vskip0.5ex
\noindent
{\it Pseudo-scalar decay constants.}
Since the calculation of the axial-current renormalization 
constant does not require separate simulations,
the renormalized pseudo-scalar meson decay constants
become directly accessible on each simulated lattice.
Depending on the flavour channel and
the precision that is to be attained,
the mass-dependent O($a$) corrections in the PCAC
relation used to determine the renormalization constant
$\ZAh{rs}$ may however need to be estimated non-perturbatively.

\vskip0.5ex
\noindent
{\it Scaling to the continuum limit.}
The results of the computations reported in this section
depend on a choice of flow times.
When lattices with different spacings are simulated,
these parameters should be held fixed
in units of a suitable reference scale such
as the reference flow time $t_0$ [\ref{WilsonFlow}].
Only then can the calculated renormalized quantities 
be expected to converge to the continuum limit
with a rate proportional to $a^2$.

\section 10. Concluding remarks

Through its extension to the quark fields,
the gradient flow becomes a powerful tool for 
studies of the spontaneous breaking of chiral symmetry in QCD
and the physics of the light pseudo-scalar mesons.
The focus in this paper has been on
the theoretical framework and its implementation in 
one of the widely used formulations of lattice QCD. 
In view of the renormalizability properties of the flow,
the choice of the lattice regularization is however expected to be
irrelevant in the continuum limit
as long as the locality and gauge invariance 
of the lattice theory are guaranteed.
The theoretical analysis obviously applies to 
field theories
with other gauge groups and fermion multiplets as well.

In the cases worked out so far,
the application of the gradient flow in numerical lattice QCD
offers important technical advantages 
with respect to the established computational strategies. 
Apart from the ones already mentioned in sect.~1, 
there are potentially many further uses of the flow.
The flavour singlet channel, for example,
has not been considered yet and the time-dependent condensate
is likely to be an easily accessible chiral
order parameter in QCD at non-zero temperatures.
Moreover, the application of the 
flow in QCD-like theories near the conformal window may 
conceivably lead to interesting qualitative insights.

\vskip1.0ex minus 0.5ex
All calculations reported in sect.~9 were performed on a 
dedicated PC cluster at CERN. I am grateful 
to the CERN management for providing the required funds
and to the CERN IT Department for technical support.

\appendix A. Notational conventions

\vskip-2.5ex

\subsection A.1 Gauge group

The Lie algebra of $\SUn$ 
may be identified with the linear space of all complex, anti-hermitian
traceless $N\times N$ matrices.
With respect to a 
basis $T^a$, $a=1,\ldots,N^2-1$, of such matrices, 
the general element $X$ of the Lie algebra is given by
$X=X^aT^a$ with real components $X^a$
(repeated group indices are automatically summed over).
The generators are assumed to satisfy
\equation{
  \tr\{T^aT^b\}=-\frac{1}{2}\delta^{ab},
  \enum
}
but are otherwise left unspecified.

Gauge fields are normalized such that
gauge transformations and covariant derivatives 
do not involve the gauge coupling.
Lorentz indices $\mu,\nu,\ldots$ in $D$ dimensions 
range from $0$ to $D-1$ and are
summed over when they occur in matching pairs.
The space-time metric is assumed to be Euclidean. In particular,
$p^2=p_{\mu}p_{\mu}$ for any momentum $p$ and $\delta_{\mu\mu}=D$.

\subsection A.2 Dirac matrices

The Dirac matrices $\dirac{\mu}$ satisfy
\equation{
  \diracstar{\mu}{\dagger}=\dirac{\mu},\qquad
  \{\dirac{\mu},\dirac{\nu}\}=2\delta_{\mu\nu}.
  \enum
}
Scalar products $\dirac{\mu}p_{\mu}$
are abbreviated by
$\slash{p}$, as usual, and 
$\sigma_{\mu\nu}={i\over2}\left[\dirac{\mu},\dirac{\nu}\right]$.
In any dimension $D$ close to 4,
\equation{
  \dirac{5}=\dirac{0}\dirac{1}\dirac{2}\dirac{3},
  \qquad
  \tr\{1\}=4.
  \enum
}
Note that the axial currents require renormalization if dimensional
regularization is used with these conventions for the Dirac matrices.

\subsection A.3 Lattice theory

The lattice theories considered in this paper are 
set up as usual on a four-dimen\-sional hypercubic 
lattice with spacing $a=1$. 
For the quark fields, Wilson's formulation is chosen,
where the quark fields reside on the sites of the lattice
and carry the same indices as the quark fields in the 
continuum theory.
The gauge covariant forward and backward difference operators
acting on a quark field $\psi(x)$ in presence of 
the gauge field $U(x,\mu)$ are defined by
\equation{
  \nab{\mu}\psi(x)=U(x,\mu)\psi(x+\hat{\mu})-\psi(x),
  \enum
  \nexteq{2.5ex}
  \nabstar{\mu}\psi(x)=\psi(x)-U(x-\hat{\mu},\mu)^{-1}\psi(x-\hat{\mu}),
  \enum
}
where $\hat{\mu}$ denotes the unit vector in direction $\mu$.
On the lattice, $\drv{\mu}$ and $\drvstar{\mu}$ stand for
the ordinary forward and backward difference operators 
and
\equation{
  \ring{\partial}_{\mu}=\frac{1}{2}(\drv{\mu}+\drvstar{\mu})
  \enum
}
for their symmetric combination.

The differential operators $\partial^a_{x,\mu}$ act on 
differentiable functions $f(U)$ of the gauge field $U$ according to
\equation{
  \partial^a_{x,\mu}f(U)=
  {\rmd\over\rmd s}f(\rme^{sX}U)\!\left.{{\vphantom{r\over s}}}\right|_{s=0},
  \hskip1.7em
  X(y,\nu)=\cases{T^a & if $(y,\nu)=(x,\mu)$,\cr
                          \noalign{\vskip0.8ex}
                        0  & otherwise.\cr}
  \enum
}
While these operators depend on the choice of the generators $T^a$,
the gradient field
\equation{
  (\partial f)(U,x,\mu)=T^a\partial^a_{x,\mu}f(U)
  \enum
}
can be shown to be basis-independent.

\appendix B. Properties of the mapping (5.9)

The mapping (5.9) is differentiable and satisfies
\equation{
  \Psu(\rme^{sX})\mathrel{\mathop=_{s\to0}}
  sX+\rmO(s^2)
  \enum
}
for all matrices $X$ in the Lie algebra of $\SUn$. 
These properties imply that
there exists an open neighbourhood of unity in $\SUn$,
where $\Psu$ is one-to-one and in which
the Jacobian matrix
\equation{
  J(U)^{ab}=-2\kern1pt\tr\kern-1pt\left\{T^a\Psu(T^bU)\right\}
  \enum
} 
is nowhere singular.

Using simple norm estimates, 
the interior of the domain
\equation{
  \Dsu=\left\{U\in\SUn\bigm|
  \Re\tr\{1-U\}\leq(16N)^{-1}\right\}
  \enum
}
can be shown to be such a neighbourhood of unity. Moreover,
the Jacobian $\det J(U)$ is strictly positive 
on $\Dsu$.
As a consequence, the identity
\equation{
  \int_{\Dsu} \rmd U\,\delta\left(\Psu(U)-\Psu(V)\right)f(U)=
  k\det J(V)^{-1}f(V)
  \enum
}
holds for all integrable functions $f(U)$
and all $V\in\Dsu$, the Dirac $\delta$-function being the 
one associated to the $\SUn$ invariant measure 
on the Lie algebra.
The proportionality 
constant $k$ in this equation is positive and could be worked out
explicitly, but its value is not needed in this paper.

\appendix C. Wick contractions

The Wick contractions of the basic fermion fields 
in the O($a$) improved lattice theory 
are first worked out for the case of the 
discretized flow equations with time step $\eps$.
Time-ordering ambiguities are avoided in this way and
the limit $\eps\to0$ can then easily be taken at the end of 
the calculation.

\subsection C.1 Fermion action and functional integral

To get started, it may be helpful to recall that
the fermion part of the total action of the O($a$) improved lattice 
theory in 4+1 dimensions is the sum of
a boundary term,
\equation{
  \SF=\sum_x\bigl\{\psibar(x)\Dm\psi(x)
  +\cfl\lambdabar(0,x)\lambda(0,x)\bigr\},
  \enum
}
and of the bulk action (5.8).
The Wilson--Dirac operator $\Dm$ in eq.~(C.1) includes the
Sheikholeslami--Wohlert term [\ref{SW},\ref{SFimp}] and the 
mass term with bare mass matrix $M_0$.

In the bulk action (5.8), it is understood that the 
time-dependent fields satisfy the boundary conditions (2.6).
The unconstrained fermion (and similarly the anti-fermion) fields 
integrated over in the functional integral can thus be
taken to be 
\equation{
  \psi(x),\quad\chi(\eps,x),\ldots,\chi(T,x),\quad
  \lambda(0,x),\ldots,\lambda(T-\eps,x).
  \enum
}
Since the action is quadratic in these fields,
the fermion integral is given through Wick's theorem.

\subsection C.2 Field transformation

In the following, an important r\^ole is played by
the kernel $K_{\eps}(t,x;s,y)$ that satisfies
\equation{
  K_{\eps}(t,x;t,y)=\delta_{xy}
  \enum
}
at all $t$ in the range $0\leq t\leq T$ and 
the discretized quark flow equation
\equation{
  \left(\partial_t-\Delta\right)K_{\eps}(t,x;s,y)=0
  \enum
}
if $0\leq s\leq t<T$.
Clearly, since the flow equation (C.4) can be solved 
step by step starting from time $t=s$,
these equations uniquely determine the kernel at
all $t\geq s$.

The calculation of the fermion propagators can now be simplified by
performing a field transformation, where 
$\chi(t,x)$ is expressed through $\psi(x)$ and 
another field $\phi(t,x)$, defined at $t=\eps,\ldots,T$,
according to
\equation{
  \chi(t,x)=\sum_yK_{\eps}(t,x;0,y)\psi(y)
  +\eps\sum_{s=\eps}^t\sum_yK_{\eps}(t,x;s,y)\phi(s,y).
  \enum
}
For any given field $\psi(x)$,
the mapping from the field variables
$\chi(\eps,x),\ldots,\chi(T,x)$ to the new variables 
$\phi(\eps,x),\ldots,\phi(T,x)$ is invertible in view of the 
identity
\equation{
  \phi(t+\eps,x)=\cases{
  \left(\partial_t-\Delta\right)\chi(t,x) & if $0<t<T$,\cr
  \noalign{\vskip1.5ex}
  {1\over\eps}(\chi(\eps,x)-\psi(x))-\Delta\psi(x) &
  if $t=0$.\cr
  }
  \enum
}
Moreover, the associated Jacobian matrix is lower triangular
and its determinant does not depend on the gauge field.

After the field transformation and the corresponding
transformation of the anti-fermion fields,
the fermion action assumes the form
\equation{
  \SF+\SFfl=\sum_x
  \bigl\{\psibar(x)\Dm\psi(x)+\cfl\lambdabar(0,x)\lambda(0,x)\bigr\}
  \noenum
  \nexteq{0.0ex}
  {\phantom{\SF+\SFfl=}}
  +\eps\sum_{t=0}^{T-\eps}\sum_x\bigl\{
  \lambdabar(t,x)\phi(t+\eps,x)+\phibar(t+\eps,x)\lambda(t,x)\bigr\}.
  \enum
}
The transformation thus achieves a nearly perfect decoupling of the 
field variables.

\subsection C.3 Contractions

Since the fundamental quark fields $\psi,\psibar$ are completely
decoupled from the other fields, the contraction of these fields 
\equation{
  \wick{\psi(x)\psibar(y)}=S(x,y),
  \enum
  \nexteq{2.5ex}
  \Dm S(x,y)=\delta_{xy},
  \enum
}
coincides with the quark propagator in the theory in 4 dimensions
as expected.
The other non-vanishing contractions deriving 
from the action (C.7) are
\equation{
  \verylongwick{\phi(t+\eps,x)\lambdabar(t,y)}=
  \longwick{\lambda(t,x)\phibar(t+\eps,y)}={1\over\eps}\delta_{xy},
  \quad t=0,\eps,\ldots T-\eps,
  \enum
  \nexteq{2.5ex}
  \longwick{\phi(\eps,x)\phibar(\eps,y)}=-{\cfl\over\eps^2}\delta_{xy}.
  \enum
}
For the contractions involving the fields $\chi$ and $\chibar$
one then obtains
\equation{
  \longwick{\chi(t,x)\lambdabar(s,y)}=
  \theta(t>s)K_{\eps}(t,x;s+\eps,y),
  \enum
  \nexteq{3.0ex}
  \longwick{\lambda(t,x)\chibar(s,y)}=
  \theta(t<s)K_{\eps}(s,y;t+\eps,x)^{\dagger},
  \enum
  \nexteq{3.0ex}
  \longwick{\chi(t,x)\psibar(y)}=\sum_{v}
  K_{\eps}(t,x;0,v)S(v,y),
  \enum
  \nexteq{2.4ex}
  \wick{\psi(x)\chibar(s,y)}=\sum_{w}
  S(x,w)K_{\eps}(s,y;0,w)^{\dagger},
  \enum
  \nexteq{2.4ex}
  \longwick{\chi(t,x)\chibar(s,y)}=\sum_{v,w}
  K_{\eps}(t,x;0,v)S(v,w)
  K_{\eps}(s,y;0,w)^{\dagger}
  \noenum
  \nexteq{1.5ex}
  {\phantom{\longwick{\chi(t,x)\chibar(s,y)}=}}
  -\cfl\sum_vK_{\eps}(t,x;\eps,v)K_{\eps}(s,y;\eps,v)^{\dagger}.
  \enum  
}
In all these equations, it is understood that the flow-time arguments of 
$\chi$ and $\chibar$ are in the range $(0,T]$ and those of 
$\lambda$ and $\lambdabar$ in the range $[0,T)$.

\subsection C.4 Contractions in the continuous time limit

When the time step $\eps$ is taken to zero, the kernel $K_{\eps}(t,x;s,y)$
smoothly converges to the fundamental solution $K(t,x;s,y)$
of the quark flow equation in the continuous time formulation of
the lattice theory.
The complete set of the non-zero contractions of the fermion
fields, given by eqs.~(C.8) and (C.12)--(C.16), therefore has a 
well-defined limit.

\appendix D. Integration of the lattice flow equations

The numerical integration of the pure-gauge gradient flow has already been 
discussed in appendix C of ref.~[\ref{WilsonFlow}]. 
Here the 3rd order Runge--Kutta integrator described there 
is extended to the quark flow equation.

\subsection D.1 Integration of the gradient flow

Following ref.~[\ref{WilsonFlow}],
the flow equation (5.2) is written
in an abstract form
\equation{
   \partial_tV_t=Z(V_t)V_t,
   \enum
}
where the gauge field $V_t$
is considered to be an element of a 
(high-dimensional) Lie group and the gradient $Z(V_t)$
of the Wilson action an element of the associated Lie algebra.
The Runge--Kutta integrator mentioned above 
proceeds in time steps of size $\eps$.
Assuming $V_t$ is known at some flow time $t$, an approximation to 
the exact solution $V_{t+\eps}$ at time $t+\eps$ 
is obtained by computing the fields
\equation{
  W_0=V_t,
  \noenum
  \nexteq{2.5ex}
  W_1=\exp\bigl\{\frac{1}{4}Z_0\bigr\}W_0,
  \noenum
  \nexteq{2.5ex}
  W_2=\exp\bigl\{\frac{8}{9}Z_1-\frac{17}{36}Z_0\bigr\}W_1,
  \noenum
  \nexteq{2.5ex}
  W_3=\exp\bigl\{\frac{3}{4}Z_2
                 -\frac{8}{9}Z_1+\frac{17}{36}Z_0\bigr\}W_2,
  \enum
}
where
\equation{
  Z_i=\eps Z(W_i),\qquad i=0,1,2.
  \enum
}
These rules are such that
\equation{
  W_3=V_{t+\eps}+\rmO(\eps^4)
  \enum
}
and $W_3$ thus provides the desired approximation to $V_{t+\eps}$.

\subsection D.2 Integration of the quark flow equation

It is again helpful to rewrite
the flow equation in an abstract form,
\equation{
  \partial_t\chi_t=\Delta(V_t)\chi_t,
  \enum
}
where the quark field $\chi_t$
is considered to be an element of a 
complex vector space. 
Given $V_t$ and $\chi_t$ at some flow time $t$, and assuming
the gauge fields $W_0$, $W_1$ and $W_2$ are as above,
the quark fields
\equation{
  \phi_0=\chi_t,
  \noenum
  \nexteq{2.5ex}
  \phi_1=\phi_0+\frac{1}{4}\Delta_0\phi_0,
  \noenum
  \nexteq{2.5ex}
  \phi_2=\phi_0+\frac{8}{9}\Delta_1\phi_1-\frac{2}{9}\Delta_0\phi_0,
  \noenum
  \nexteq{2.5ex}
  \phi_3=\phi_1+\frac{3}{4}\Delta_2\phi_2,
  \enum
}
may then be calculated, where
\equation{
  \Delta_i=\eps\Delta(W_i),\qquad i=0,1,2.
  \enum
}
It is not difficult to show that
\equation{
  \phi_3=\chi_{t+\eps}+\rmO(\eps^4),
  \enum
}
and $\phi_3$ thus approximates $\chi_{t+\eps}$ 
to an accuracy that matches the one attained by 
the integrator of the pure-gauge flow.

\appendix E. Integration of the adjoint flow equation (7.10)

The numerical integration of the adjoint flow equation
is complicated by the fact that the 
backward integration of the flow equation for the gauge field
is exponentially unstable. 
In this appendix, the problem is first reformulated
and its solution is then discussed. For simplicity, the index
$k$ in eq.~(7.10) is omitted.

\subsection E.1 Adjoint Runge--Kutta integration

For a given step size
$\eps>0$ and $t=n\eps$, $n=0,1,2,\ldots$, let
\equation{
  V_t^{\eps}(x,\mu)\quad\hbox{and}\quad
  \chi_t^{\eps}(x)
  \enum
}
be the gauge and quark fields generated by the Runge--Kutta integrator
defined in appendix D. There exists another sequence of quark fields
\equation{
  \xi_s^{\eps}(x),\quad s=t,t-\eps,t-2\eps,\ldots,0,
  \enum
}
with initial value
\equation{
  \xi_t^{\eps}(x)=\eta(x),
  \enum
}
such that the identity
\equation{
  (\xi_s^{\eps},\chi_s^{\eps})=(\eta,\chi_t^{\eps})
  \enum
}
holds exactly for all $s$. This property ensures that
$\xi_s^{\eps}$ is an approximate solution of the adjoint
flow equation, accurate to order $\eps^3$, 
with the required initial value.

The Runge--Kutta steps that lead from $\chi_s^{\eps}$
to $\chi_{s+\eps}^{\eps}$ amount to applying the operator
\equation{
  R_s^{\eps}=1+\frac{1}{4}\Delta_0+\frac{3}{4}\Delta_2
  \bigl\{1-\frac{2}{9}\Delta_0+\frac{8}{9}\Delta_1(1+\frac{1}{4}\Delta_0)
  \bigr\}
  \enum
}
to $\chi_s^{\eps}$, where $\Delta_i$ is given by eq.~(D.7) 
and the fields $W_i$ that appear there 
by eq.~(D.2) with $W_0=V_s^{\eps}$.
From eq.~(E.4) one then infers that
\equation{
  \xi_s^{\eps}(x)=({R_s^{\eps}}^{\dagger}\xi_{s+\eps}^{\eps})(x)
  \enum
  \nexteq{2.5ex}
  {R_s^{\eps}}^{\dagger}=1+\frac{1}{4}\Delta_0+
  \bigl\{1-\frac{2}{9}\Delta_0+(1+\frac{1}{4}\Delta_0)\frac{8}{9}\Delta_1
  \bigr\}\frac{3}{4}\Delta_2,
  \enum
}
and $\xi_s^{\eps}$ is thus obtained from 
$\xi_{s+\eps}^{\eps}$ by calculating
\equation{
  \lambda_3=\xi_{s+\eps}^{\eps},
  \noenum
  \nexteq{2.5ex}
  \lambda_2=\frac{3}{4}\Delta_2\lambda_3,
  \noenum
  \nexteq{2.5ex}
  \lambda_1=\lambda_3+\frac{8}{9}\Delta_1\lambda_2,
  \noenum
  \nexteq{2.5ex}
  \lambda_0=\lambda_1+\lambda_2+
  \frac{1}{4}\Delta_0(\lambda_1-\frac{8}{9}\lambda_2),
  \enum
}
and setting $\xi_s^{\eps}=\lambda_0$.

\subsection E.2 Stability issue

The adjoint recursion (E.8) is a smoothing operation and is therefore
numerically stable. To be able to apply the recursion at time $s+\eps$,
the gauge field $V_s^{\eps}$ at flow time $s$ must however be known.
In principle, the sequence of gauge fields from $s=t$ to $s=0$
could be reconstructed through
backward integration of the gradient flow, 
but such strategies are not practical,
because the flow is exponentially unstable in this direction.

Alternatively, $V_s^{\eps}$ can be computed through
forward integration of the flow starting from the fundamental
field $V_0^{\eps}=U$. If the field is recomputed for every $s$,
the total number of gauge update steps required in the 
course of the integration
scales like the square of the number $m=t/\eps$ of time steps.
This method is numerically safe and the calculated
quark fields satisfy the identity (E.4) practically to machine precision.
The required computational effort can however be prohibitively large.

\subsection E.3 Improved scheme

The adjoint flow equation can be integrated more efficiently
by dividing the sequence $0,\eps,\ldots,t$ of flow
times into $n_b$ consecutive blocks of about equal size.
One may then first compute the field $V_r^{\eps}$ at 
the time $r$ at the beginning of the last block 
by forward integration from time 0 and
keep this field in the memory of the computer. After that
the adjoint flow equation is integrated from time $t$ to $r$
by applying the update step (E.8) and 
using the safe reconstruction method described above
for the gauge field,
where the reconstruction now starts 
from the stored field rather than the field
at time $0$.

In the next step the gauge field is calculated
at the first time $r$ in the next-to-last block 
and stored in memory. The integration of the 
adjoint flow equation can then proceed to time $r$
as before. Clearly, the blocks can 
be processed in this way one after another
until the integration reaches time $0$.

The total number of gauge update steps
in this scheme is minimized if $m=n_b^2$.
While $m$ will rarely be the square of an integer,
one can always set $n_b=\lceil m^{1/2}\rceil$ and divide
$m$ into blocks of size $m_b=\lfloor m/n_b\rfloor$ and $m_b+1$.
The computational effort then scales approximately like
$m^{3/2}$ and is much smaller than the one required
for the safe scheme described in the previous subsection
already at low values of $m$.

\subsection E.4 Hierarchical scheme

A further acceleration of the computation can be achieved 
by dividing each block
into $n_b$ smaller blocks, and these into $n_b$ even smaller
blocks, and so on, with a
saved gauge field at each block level.
Given $m$ and the number $n_s$ of saved gauge fields, 
a nearly optimal choice of the block number $n_b$ is
\equation{
  n_b=\lceil m^{1/(n_s+1)}\rceil.
  \enum
}
For $m\leq10^3$ and $n_s=4$, for example, the
total number of gauge-field updates that need to be 
performed is then not more than about $5m$.
Moreover, 
when several quark fields are evolved simultaneously,
the intermediate gauge fields need to be computed only once.

\beginbibliography


\bibitem{WilsonFlow}
M. L\"uscher,
{\it Properties and uses of the Wilson flow in lattice QCD},
JHEP 1008 (2010) 071


\bibitem{RenFlow}
M. L\"uscher, P. Weisz,
{\it Perturbative analysis of the gradient flow in non-Abelian gauge
theories},
JHEP 1102 (2011) 051


\bibitem{StepScaling}
M. L\"uscher, P. Weisz, U. Wolff,
{\it A numerical method to compute the running coupling in asymptotically 
free theories},
Nucl. Phys. B359 (1991) 221

\bibitem{NogradiEtAl}
Z. Fodor, K. Holland, J. Kuti, D. Nogradi, C. H. Wong,
{\it The Yang-Mills gradient flow in finite volume},
JHEP 1211 (2012) 007; {\it The gradient flow running coupling scheme},
PoS (Lattice 2012) 050

\bibitem{FritzschRamos}
P. Fritzsch, A. Ramos,
{\it The gradient flow coupling in the Schr\"odinger functional},
arXiv:1301.4388


\bibitem{Guesken}
S. G\"usken,
{\it A study of smearing techniques for hadron correlation functions},
Nucl. Phys. B (Proc. Suppl.) 17 (1990) 361

\bibitem{AlexandrouEtAl}
C. Alexandrou, F. Jegerlehner, S. G\"usken, K. Schilling, R. Sommer,
{\it B meson properties from lattice QCD},
Phys. Lett. B256 (1991) 60


\bibitem{Zinn}
J. Zinn--Justin,
{\it Renormalization and stochastic quantization},
Nucl. Phys. B275 [FS17] (1986) 135

\bibitem{ZinnZwanziger}
J. Zinn--Justin, D. Zwanziger, 
{\it Ward identities for the stochastic quantization of gauge fields},
Nucl. Phys. B295 [FS21] (1988) 297


\bibitem{Symanzik}
K. Symanzik,
{\it Schr\"odinger representation and Casimir effect in
renormalizable quantum field theory},
Nucl. Phys. B190 [FS3] (1981) 1


\bibitem{Wilson}
K. G. Wilson, {\it Confinement of quarks}, Phys. Rev. D10 (1974) 2445



\bibitem{GinspargWilson}
P. H. Ginsparg, K. G. Wilson,
{\it A remnant of chiral symmetry on the lattice},
Phys. Rev. D25 (1982) 2649


\bibitem{Kaplan}
D. B. Kaplan,
{\it A method for simulating chiral fermions on the lattice},
Phys. Lett. B288 (1992) 342;
{\it Chiral fermions on the lattice},
Nucl. Phys. B (Proc. Suppl.) 30 (1993) 597

\bibitem{Shamir}
Y. Shamir,
{\it Chiral fermions from lattice boundaries},
Nucl. Phys. B406 (1993) 90

\bibitem{FurmanShamir}
V. Furman, Y. Shamir,
{\it Axial symmetries in lattice QCD with Kaplan fermions},
Nucl. Phys. B439 (1995) 54


\bibitem{Hasenfratz}
P. Hasenfratz, {\it Prospects for perfect actions},
Nucl. Phys. B (Proc. Suppl.) 63 (1998) 53;
{\it Lattice QCD without tuning, mixing and current renormalization},
Nucl. Phys. B525 (1998) 401

\bibitem{HLN}
P. Hasenfratz, V. Laliena, F. Niedermayer,
{\it The index theorem in QCD with a finite cutoff},
Phys. Lett. B427 (1998) 125


\bibitem{NeubergerDirac}
H. Neuberger,
{\it Exactly massless quarks on the lattice},
Phys. Lett. B417 (1998) 141;
{\it More about exactly massless quarks on the lattice},
{\it ibid.}\/ B427 (1998) 353


\bibitem{ExactChSy}
M. L\"uscher,
{\it Exact chiral symmetry on the lattice and the Ginsparg--Wilson relation},
Phys. Lett. B428 (1998) 342

\bibitem{BoulderReview}
F. Niedermayer,
{\it Exact chiral symmetry, topological charge and related topics},
Nucl. Phys. B (Proc. Suppl.) 73 (1999) 105


\bibitem{ChPT}
J. Gasser, H. Leutwyler,
{\it Chiral perturbation theory to one loop},
Ann. Phys. 158 (1984) 142


\bibitem{ArnoldI}
V. I. Arnold,
{\it Ordinary Differential Equations}, 3rd ed. (Springer-Verlag, Berlin, 2008)


\bibitem{Stout}
C. Morningstar, M. Peardon, 
{\it Analytic smearing of SU(3) link variables in lattice QCD},
Phys. Rev. D69 (2004) 054501


\bibitem{SymanzikSeffI}
K. Symanzik, 
{\it Some topics in quantum field theory},
in: Mathematical problems in theoretical physics, eds. R. Schrader
et al., Lecture Notes in Physics, Vol. 153 (Springer, New York, 1982)

\bibitem{SymanzikSeffII}
K. Symanzik,
{\it Continuum limit and improved action in lattice theories:
(I). Principles and $\varphi^4$ theory},
Nucl. Phys. B226 (1983) 187


\bibitem{OnShell}
M. L\"uscher, P. Weisz,
{\it On-shell improved lattice gauge theories},
Commun. Math. Phys. 97 (1985) 59 [E: {\it ibid.} 98 (1985) 433]


\bibitem{SW}
B. Sheikholeslami, R. Wohlert, 
{\it Improved continuum limit lattice action for QCD with Wilson fermions},
Nucl. Phys. B259 (1985) 572

\bibitem{SFimp}
M. L\"uscher, S. Sint, R. Sommer, P. Weisz,
{\it Chiral symmetry and O(a) improvement in lattice QCD},
Nucl. Phys. B478 (1996) 365


\bibitem{ImpNonD}
T. Bhattacharya, R. Gupta, W. Lee, S. R. Sharpe, J. M. S. Wu,
{\it Improved bilinears in lattice QCD with nondegenerate quarks},
Phys. Rev. D73 (2006) 034504


\bibitem{MichaelPeisa}
C. Michael, J. Peisa,
{\it Maximal variance reduction for stochastic propagators with 
applications to the static quark spectrum},
Phys. Rev. D58 (1998) 034506


\bibitem{TMopenQCD}
M. L\"uscher, S. Schaefer,
{\it  	
Lattice QCD with open boundary conditions and twisted-mass reweighting},
Comput. Phys. Commun. 184 (2013) 519


\bibitem{openQCD}
M. L\"uscher, S. Schaefer,
{\it Lattice QCD without topology barriers},
JHEP 1107 (2011) 036


\bibitem{AokiEtAlI}
S. Aoki et al.~(PACS-CS collab.), 
{\it 2+1 flavor lattice QCD toward the physical point},
Phys. Rev. D79 (2009) 034503

\bibitem{AokiEtAlII}
S. Aoki et al.~(PACS-CS collab.),
{\it Physical point simulation in 2+1 flavor lattice QCD},
Phys. Rev. D81 (2010) 074503


\bibitem{Iwasaki}
Y. Iwasaki,
{\it Renormalization group analysis of lattice theories and
improved lattice action. II -- Four-dimensional non-Abelian SU(N) gauge model},
preprint UTHEP-118 (1983) and arXiv:1111.7054


\bibitem{AokiEtAlIII}
S. Aoki et al. (CP-PACS collab.),
{\it Nonperturbative O(a) improvement of the Wilson quark action with the 
RG-improved gauge action using the Schr\"odinger functional method},
Phys. Rev. D73 (2006) 034501

\bibitem{KanekoEtAl}
T. Kaneko et al. (CP-PACS/JLQCD and ALPHA collab.),
{\it Non-perturbative improvement of the axial current with 
three dynamical flavors and the Iwasaki gauge action},
JHEP 0704 (2007) 092


\bibitem{AokiEtAlIV}
S. Aoki et al. (PACS-CS collab.),
{\it Non-perturbative renormalization of quark mass 
in $N_f$=2+1 QCD with the Schr\"odinger functional scheme},
JHEP 1008 (2010) 101


\bibitem{ZASFI}
M. L\"uscher, S. Sint, R. Sommer, H. Wittig,
{\it Non-perturbative determination of the axial current 
normalization constant in O(a) improved lattice QCD},
Nucl. Phys. B491 (1997) 344

\bibitem{ZASFII}
M. Della Morte, R. Hoffmann, F. Knechtli, R. Sommer, U. Wolff,
{\it Non-pertur\-ba\-tive renormalization of the axial current 
with dynamical Wilson fermions},
JHEP 0507 (2005) 007

\endbibliography

\bye